\documentclass[final]{elsarticle}

\usepackage{hyperref}

\usepackage[table]{xcolor}

\journal{Journal of \LaTeX\ Templates}

\usepackage{epsfig}
\usepackage{graphicx}
\usepackage{amssymb}
\usepackage{algorithm}
\usepackage[noend]{algpseudocode}
\usepackage{multirow}
\usepackage{subfigure}
\usepackage{balance}
\usepackage{sidecap}

\usepackage{color}
\usepackage{listings}
\begin{document}

\begin{frontmatter}

\title{Efficient Methods and Parallel Execution for Algorithm Sensitivity Analysis with Parameter Tuning on Microscopy Imaging Datasets}

\author{George Teodoro\fnref{1}, Tahsin Kurc, Lu\'is F. R. Taveira, Alba C. M. A. Melo, Jun Kong, and Joel Saltz}
\address{Department of Computer Science, University of Bras\'ilia, Bras\'ilia, 70910-900, Brazil}
\address{Biomedical Informatics Department, Emory University, Atlanta, 30322, USA}
\address{Biomedical Informatics Department, Stony Brook University, Stony Brook, 11794-8322, USA}
\address{Scientific Data Group, Oak Ridge National Laboratory, Oak Ridge, USA}

\maketitle

\begin{abstract}
\textbf{Background:} We describe an informatics framework for researchers and clinical 
investigators to efficiently perform parameter sensitivity analysis and auto-tuning 
for algorithms that segment and classify image features in a large dataset of 
high-resolution images. The computational cost of the sensitivity analysis process can be 
very high, because the process requires processing the input dataset several times to systemtically 
evaluate how output varies when input parameters are varied. Thus, high performance computing 
techniques are required to quickly execute the sensitivity analysis process.

\noindent\textbf{Results:} We carried out an empirical evaluation of 
the proposed method on high performance computing clusters with multi-core 
CPUs and co-processors (GPUs and Intel Xeon Phis). Our results show that 
(1) the framework 
achieves excellent scalability and efficiency on a high performance computing 
cluster -- execution efficiency remained above 85\% in all experiments; (2) 
the parameter auto-tuning methods are able to
converge by visiting only a small fraction (0.0009\%) of the search space with
limited impact to the algorithm output (0.56\% on average). 

\noindent\textbf{Conclusions:} The
sensitivity analysis framework provides a range of strategies for the efficient 
exploration of the parameter space, as well as multiple indexes
to evaluate the effect of parameter modification to outputs or even correlation
between parameters. Our work demonstrates the feasibility of performing 
sensitivity analyses, parameter studies, and
auto-tuning with large datasets with the use of
high-performance systems and techniques. The proposed techologies will enable
the quatification of error estimations and output variations in these
pipelines, which may be used in application specific ways to assess uncertainty
of conclusions extracted from data generated by these image analysis pipelines.

\end{abstract}

\begin{keyword}
\texttt{Microscopy Imaging \sep Sensitivity Studies \sep Auto-tuning}
\end{keyword}

\end{frontmatter}

\section{Background} \label{sec:intro}
We define algorithm sensitivity analysis as the process of comparing results from 
multiple analyses of a dataset using variations of an analysis workflow (e.g., 
different input parameters or different algorithm versions) and quantifying 
differences in the results. This process is executed in
many phases of scientific research, including validation of models, parameter
studies and error estimation. In validation and error estimation, comparison of
multiple analyses can be used to quantify and evaluate how much models differ
or agree, and differences among output from different models can be combined in
application specific ways to assess errors and uncertainty in output.
Application parameter tuning is another important task in sensitivity analysis. In
this task, the parameter space of an analysis workflow is searched by comparing 
analysis results with {\em ground truth} to find the set of parameters which 
produces results that are closest to the ground truth with respect to some 
comparison metric. 

\subsection{Motivation} 
The primary motivation for our work is to support large scale biomedical imaging studies, 
in particular those working with large numbers of whole slide tissue images. High-resolution 
microscopy imaging of tissue specimens enables the study of disease at the cellular
and sub-cellular levels. Characterizing the sub-cellular morphology of tissue can 
lead to a better understanding of disease mechanisms and a better assessment 
of response to treatment. It is increasingly becoming feasible, with wider availability 
of advanced tissue scanners at lower price points, for research studies to collect 
tens of thousands of high resolution images. A contemporary digital microscopy scanner can 
capture a whole slide tissue image containing 20 billion pixels (using 40X objective 
magnification) in a few minutes.  A scanner with a slide loader and auto-focusing mechanism 
can generate hundreds of images in one or two days~\cite{kurc2015scalable}. 

\begin{table*}[htb!]
\begin{center}
\vspace*{-2ex}
\caption{Parameters their value ranges for two example workflows.}
\begin{scriptsize}
\subtable[Parameters of the Watershed based segmentation workflow. The search space of the segmentation contains about 21 trillion points.]{
\begin{tabular}{l l l l }
\hline
Parameter		& Description   						& Range Value					\\ \hline
Target Image 		& Image used as a normalization target 				& TI $\in [Img1, Img2,...,Img4]$ 		\\ \hline
B/G/R  			& Background detection thresholds 				&  B, G, R $\in [210, 220,...,240]$ 		\\ \hline
T1/T2			& Red blood cell detection thresholds				& T1,T2 $\in [2.5,3.0,...,7.5]$			\\ \hline
\multirow{2}{*}{G1/G2}	& \multirow{2}{*}{Thresholds to identify candidate nuclei}	& G1 $\in [5,10,...,80]$			\\ 
			& 								& G2 $\in [2,4,...,40]$				\\ \hline
\multirow{2}{*}{MinSize}& Filter out objects with area (pixels) $<$			& \multirow{2}{*}{MinSize $\in [2,4,...,40]$} 	\\ 
			& MinSize after candidate nuclei identification  		&  						\\ \hline
\multirow{2}{*}{MaxSize}& Filter out objects with area larger than 			& \multirow{2}{*}{MaxSize $\in [900,950,...,1500]$}\\ 
			& MaxSize after candidate nuclei identification 	 	&  						\\ \hline
\multirow{2}{*}{MinSizePl}& Filter out objects with area smaller than 			& \multirow{2}{*}{MinSizePl $\in [5,10,...,80]$}\\
			&  MinSizePl before watershed is executed			& 						\\ \hline
\multirow{2}{*}{MinSizeSeg}& Filter out nuclei with area smaller than			& \multirow{2}{*}{MinSizeSeg $\in [2,4,...,40]$} \\ 
			& MinSizeSeg from final segmentation 				&  						\\ \hline
\multirow{2}{*}{MaxSizeSeg}& Filter out nuclei with area smaller than 			& \multirow{2}{*}{MaxSizeSeg $\in [900,950,...,1500]$}	\\ 
			& MaxSizeSeg from final segmentation 				& 						\\ \hline
FillHoles Structure	& Structure of the propagation neighborhood			& FillHoles $\in [4$-conn$,8$-conn$]$		\\ \hline
MorphRecon Structure	& Structure of the propagation neighborhood			& MorphRecon $\in [4$-conn$,8$-conn$]$		\\ \hline
Watershed Structure	& Structure of the propagation neighborhood			& Watershed $\in [4$-conn$,8$-conn$]$		\\ \hline
\end{tabular}
}
\subtable[Parameters of the Level Set based segmentation workflow. Dummy parameter is used here to account for application output variability
due to its stochastic behaviors. The search space of the segmentation, excluding the dummy parameter contains to 2.8 billion points.]{
\begin{tabular}{l l l l }
\hline
Parameter	& Description   					& Range Value				\\ \hline
Target Image 	& Image used as a normalization target 			& TI $\in [Img1, Img2,...,Img4]$ 	\\ \hline
OTSU 		& Weighting value applied to the OTSU threshold value	& OTSU $\in [0.3,0.2,...,1.3]$ 		\\ \hline
Curvature Weight& Curvature weight (CW) of the level set functions	& CW $\in [0.0,0.05,...,1.0]$		\\ \hline
MinSize		& Minimum object size in micron	per dimension 		& MinSize $\in [1,2,...,20]$ 		\\ \hline
MaxSize		& Maximum object size in micron	per dimension		& MaxSize $\in [50,55,...,400]$ 	\\ \hline
MsKernel	& Spatial radius used in the Mean-Shift calculation	& MsKernel $\in [5,6,...,30]$		\\ \hline
LevetSetIt	& Number of iterations of the level set computation	& LevetSetIt $\in [5,6,...,150]$	\\ \hline 
\end{tabular}
}
\end{scriptsize}
\label{tab:parameters}
\end{center}
\end{table*}

\begin{figure*}[htb!]
\begin{center}
	\subfigure[fig:water-app][Watershed based segmentation workflow.]{
	\includegraphics[width=\textwidth]{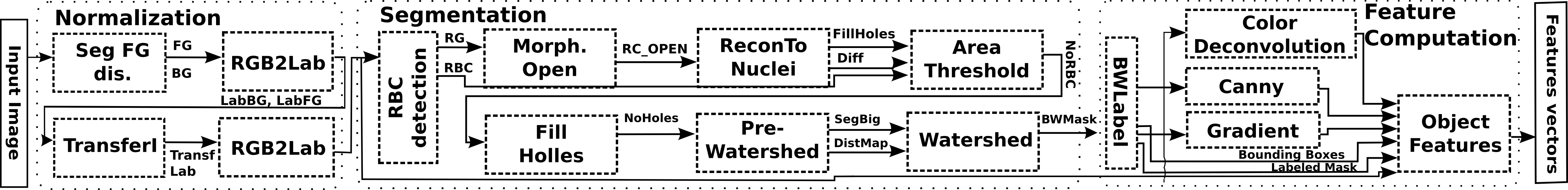}}
	\subfigure[fig:level-app][Level Set based segmentation workflow.]{
	\includegraphics[width=\textwidth]{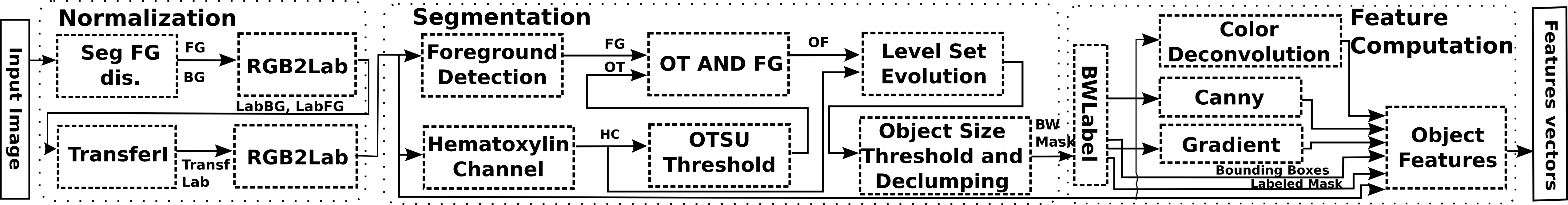}}

\caption{Two example analysis workflows with their cascade of internal
operations. They have the same high level computation stages: normalization,
segmentation and feature computation. The stages of normalization and
feature computation are shared among between the workflows, but different techniques
are used to implement the segmentation phase. The first
(Figure~(a)) uses morphological operations along with watershed
to delineate and separate clumped cells. The second
(Figure~(b)) employs the level set strategy and a mean shift
based clustering to delineate and separate clumped cells}
\label{fig:example-app}
\end{center}
\end{figure*}

Analysis of tissue images involves extraction of salient information from the
images in the form of segmented objects (e.g., nuclei or cells) and their size,
shape, intensity and texture features. These features are then used to develop
morphological models of the specimens under study to gain new insights.  A
typical analysis workflow consists of normalization, segmentation, feature
computation, feature refinement and classification operations.  The first three
analysis stages  are typically the most computationally expensive phases of an
analysis run. Figure~\ref{fig:example-app} presents two analysis workflows used
in this work. These workflows share the same high level structure, but
implement the segmentation stage using different strategies. The first
(Figure~\ref{fig:example-app}) uses morphological operations and watershed in
the segmentation~\cite{kong2013machine}, whereas the second
(Figure~\ref{fig:example-app}) performs the segmentation based on level set and
mean shift clustering~\cite{doi:10.1117/12.2217029,gomes2015efficient}. The cascade of operations
employed by each of the workflows are presented in the figures.

\begin{figure*}[htb!]
\begin{center}
\includegraphics[width=0.70\textwidth]{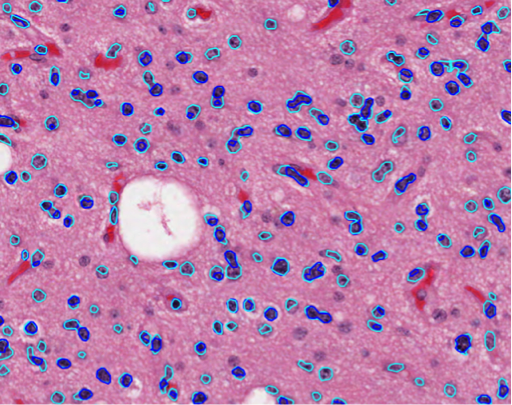}
\caption{Nuclear segmentation results from two analysis runs.}
\label{fig:two-runs}
\end{center}
\end{figure*}

Most image analysis workflows are sensitive to variations in input parameters.
A workflow optimized for a group of images or a type of tissue may not perform
well for other tissue types or images. For instance, accurate segmentation of
cancer nuclei is an important image analysis task in digital pathology.
Boundaries of segmented nuclei are algorithm dependent.  Input parameters, such
as the choice of methods for seed selection, intensity thresholds for boundary
detection and thresholds for separation of clumped nuclei, will impact the
results (the number and locations of detected nuclei, the shape of boundaries
of a nucleus, etc). Figure~\ref{fig:two-runs} shows nuclear segmentation
results from two analysis runs. The two analysis pipelines have good agreement
in some regions (i.e., the boundaries of the polygons overlap closely) and
large disagreement in other regions, where either one algorithm has not
segmented nuclei while the other has or there are large differences between the
boundaries of a nucleus segmented by the two algorithms. 

It is, therefore, important to quantify the impact of changes 
in input parameters to output generated by different stages and by the overall workflow.
In this work, we focus on the segmentation stage as our example, because this is a 
crucial stage in extracting morphological information from images and consists of 
several complex and parameterized data transformation steps. An approach
for evaluating sensitivity of an analysis workflow and tuning its parameters is for the user
to manually run the workflow on small patches of images, visually evaluate the
results, modify the workflow parameters, and repeat the process until a
satisfactory set of parameters is
obtained~\cite{Torsney-Weir:2011:TPP:2068462.2068683,journals/tvcg/SchultzK13}.

\subsection{Data and Computation Challenges}
This labor-intensive and error-prone process does not scale to  multiple
workflows and large sets of images.  About 400,000 nuclei are identified in a
WSI, and this segmentation process takes at least an hour on a single CPU-core.
A parameter analysis run may require the evaluation of hundreds or thousands
points (parameter combinations), and each of them correspond to a full run of
the analyses pipelines on the input data. As such, for instance, the execution
of a sensitivity analysis study involving the evaluation of 2,000 points in the
application parameter space (application run) with 55~WSIs (The Cancer Genome
Atlas (TCGA) contains over 30,000 WSIs) would take about 18~years in a single
CPU-core and would read/stage close to 830~TB of data.

It is, thus, challenging to carry out parameter sensitivity analysis and
auto-tuning, even for the segmentation stage, because (i) the number of
possible parameter combinations of an analysis workflow is staggeringly high.
Table~\ref{tab:parameters} presents the parameters of interest in the
segmentation stage for our two use case image analysis workflows.  The space of
parameter values only for the segmentation stage for the watershed and level
set based workflows contain, respectively, over 21~trillion and 2.8~billion
points; (ii) the execution of a single parameter combination may be
computationally very demanding depending on the input dataset.  This is
critical in the context of whole slide tissue image analysis, because the
execution of a single analysis run (i.e., the evaluation of a single point in
the parameter space) can take hours or days depending on the size of the input
dataset. 

\subsection{Contributions of Our Work} 
The data and computation challenges of image analyses and sensitivity studies are 
major roadblocks to taking full advantage of advanced imaging technologies and tissue 
specimens. To make sensitivity analyses feasible, it is
necessary: (i)~to propose and utilize effective sensitivity analysis and
auto-tuning methods; and (ii)~to leverage high performance computing techniques
to accelerate these analyses. Our approach addresses these challenges by a framework that integrates the 
following components: 

\textit{Efficient Sensitivity Analysis (SA) Methods:} Sensitivity analysis
methods and techniques allow for the user to understand and quantify variability
in the output of a model, and apportion those variabilities to source of
uncertainty in the input parameters. It may be useful in several tasks, which
include understanding the output ranges, remove parameters that have little
influence from other studies etc. In this work, we propose to use a number of
global SA techniques to study sensitivity analysis in the output of microscopy
image analysis applications. They include Morris One-At-A-Time design
(MOAT)~\cite{10.2307/1269043,Campolongo20071509,Bertrand-2015,Alam:2004:UMR:1161734.1161926},
design experiments with pseudo-random parameter probe to calculate importance
metrics~\cite{saltelli2004sensitivity,10.2307/1268522}, such as Pearson and
Spearman's correlation coefficients, and Variance-based Decomposition
(VBD)~\cite{Weirs2012157,Sobol2001271}. 

\textit{Efficient Parameter Auto-tuning Methods:} We also study input parameter
tuning for the same class of applications.  In this task the input parameters
of an analysis application are systematically varied, and analysis results from
input parameter sets are compared to a reference result.  Comparison outcome is
used to adjust the parameter set and repeat the process.  The goal is to search
the application parameter space and find parameters that produce good results
with respect to a domain specific metric. 

\textit{On-the-fly Spatial Comparative Analysis Tools.} This component of our
solution deals with the computation of metrics of interest that involve the
comparison of spatial objects from segmented images. It provides a query-based
interface to the user, and is built on top of a set of core operations such as
cross-matching, overlay of objects, spatial proximity computations between
objects, and global spatial pattern discoveries. These core operations are
combined to implement metrics that include Dice Coefficient, Jaccard
Coefficient, Intersection Overlapping Area, Non-Overlapping Area (Number of
pixels differently segmented)~\cite{TA2015}.  K-Nearest Neighbor queries are
also support, and can be use to select a subset of objects of interest (e.g.,
close to region of study as a vein), which may be further passed for computing
the other quantitative metrics. This functionality is built on our runtime
system to allow for those metrics to be computed on-the-fly as objects are
identified by segmentation algorithms. It allows for an efficient execution as
the entire process is computed online without having to go through the
expensive step of loading the data in a spatial database~\cite{aji2013hadoop}.

\textit{Scalable Runtime for Execution of Sensitivity Analysis and Auto-tuning
Processes.} We have developed a framework that aims to address the processing
and data management challenges of the image analysis pipelines execution
through runtime optimizations targeted at distributed memory systems with
hybrid multi-core CPU and co-processors and multiple levels of storage.  In
this part of the work, we have optimized the analysis pipelines by
(1)~parallelizing core operations of image analysis for CPU and the Intel Xeon
Phi coprocessors~\cite{Teodoro:2014:CPA:2650283.2650645} and (2)~developing
scheduling strategies that can appropriately map computations to heterogeneous
processors. We also proposed optimizations targeting specific aspects of
parameter analysis and auto-tuning studies. They include efficient data
movement and staging, data-aware assignment of stages and operations in an
analysis workflow that exploits that optimize the system for repeated execution
of workflows with different parameters that process the same datasets.  Also,
we propose an approach to reduce the computation costs in these analyses
through the simultaneous parameter evaluation in order to eliminate common
computations in the execution of an application with multiple parameter sets.

\section{Materials and Methods}
\begin{figure}[htpb] 
\begin{center}
\includegraphics[width=1.0\textwidth]{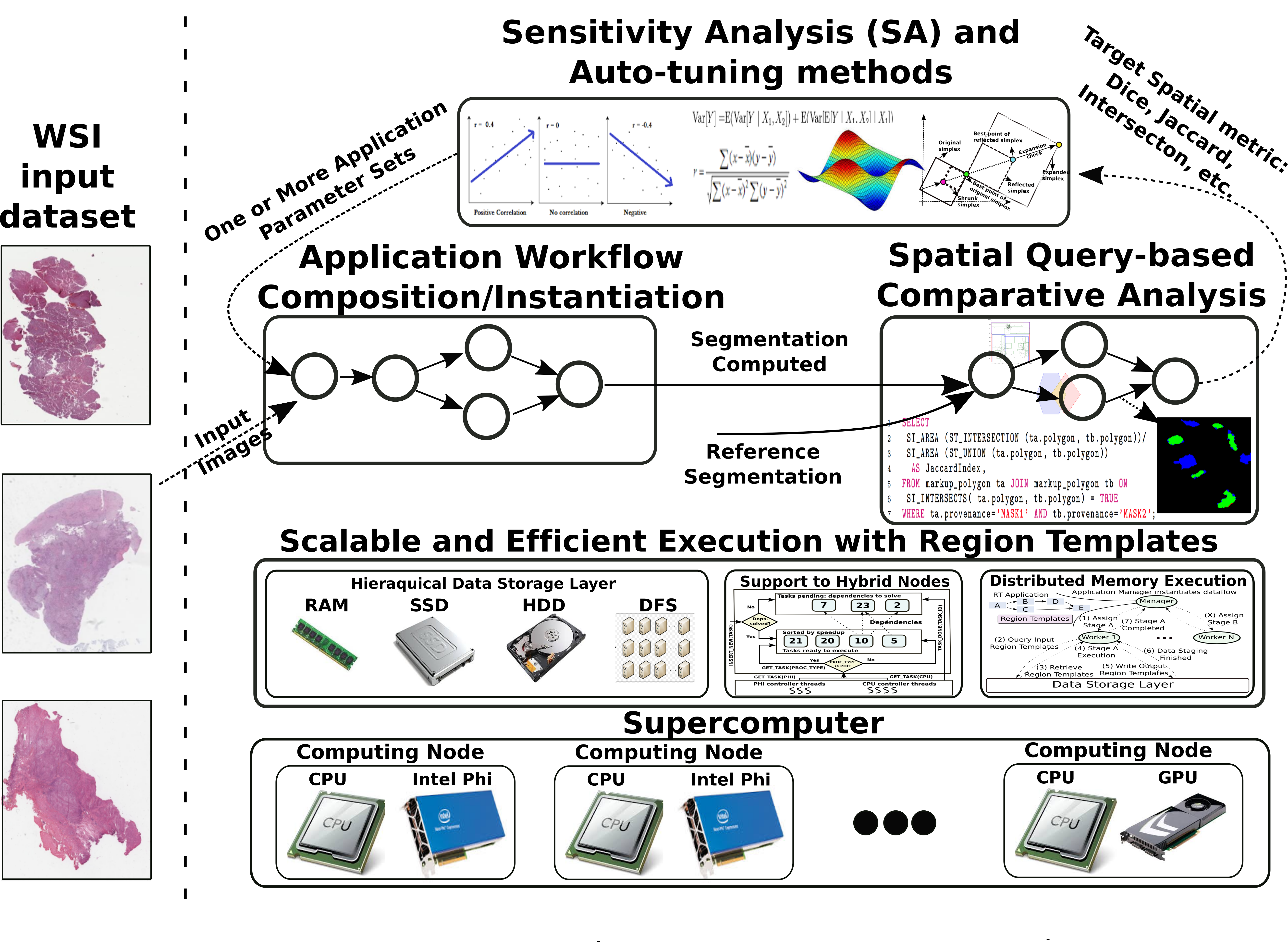} 
\caption{Overview of functionalities integration for efficient execution of
parameter sensitivity analysis and auto-tuning on large-scale imaging datasets.
A parameter study and method of interest is selected by the investigator and
application parameters are output to be evaluated.  Starting for the WSI
dataset, the application is executed on a parallel machine to compute a mask
set, which is compared to a reference mask using a query-based spatial
comparison. The comparison outputs a metric that is sent to the method
selected by the user. This cycle continues until the method has converged in
case of an optimization problem or collected enough results in a sampling based
sensitivity analysis method, for instance.} 
\label{fig:overview} 
\end{center} 
\end{figure}

%
%
%
Figure~\ref{fig:overview} shows how sensitivity analysis and parameter auto-tuning 
processes are carried out. The user specifies the set of input images, an image 
analysis workflow, input parameters to the image analysis workflow, and the metric 
of interest (e.g., Dice) for comparison of image analysis results. The image analysis 
workflow is executed using a scalable runtime environment, called Region Templates, 
on a high performance machine, while the input parameters are systematically varied 
by sensitivity analysis and auto-tuning methods supported by our framework. When image 
analysis results, i.e., sets of segmented objects, are produced, the results are 
processed by the spatial comparative analysis component. The segmentation results 
are compared to a reference segmentation dataset. The reference segmentation dataset 
may, for instance, be a previously computed set of results from earlier sensitivity 
analysis runs or a set of manual segmentation results, if the goal of sensitivity 
analysis is to tune analysis parameters to match the manual segmentation results. 
The comparison metric value generated by the comparative analysis component is then 
used as an error metric by the sensitivity analysis and auto-tuning methods which 
compute the next set of input parameters. This process is repeated until the required 
set of parameters is covered (in the case of sensitivity analysis) or error between 
the analysis results and the reference dataset is below a threshold (in the case of 
parameter auto-tuning). 
\vspace*{-1ex}
 

\subsection{Sensitivity Analysis (SA) Methods} \label{sec:sa}
 
Sensitivity Analysis (SA) is may be local or global, but in either case the
methods evaluate the variance in the output with respect to input parameter
changes.  Local analysis focuses on measuring sensitivity in the neighborhood of
a specific point in the parameter space.  Global SA, which is implemented in
this work, examines sensitivity output from the perspective of whole range of
parameter variations~\cite{Bertrand-2015}. A SA study includes selecting
uncertainty input factors and the application output of interest, running the
application with a number of points defined by a SA method, and the actual
calculation the sensitivity statistics. Several SA methods can be employed in a
study. 
%
SA methods fall into one of two types~\cite{Bertrand-2015}: methods that are
used for an initial fast exploration and are used in practice to quickly {\em
screen} non-influential input parameters that may be removed from further
analysis; and methods that compute {\em measures of importance} or quantitative
sensitivity indices. We implement methods from these two types in our framework
as described below.

\subsubsection{Methods to Screen Input Parameters}
Our current implementation supports one of the most commonly used screening
methods, called Morris One-At-A-Time (MOAT) design~\cite{10.2307/1269043}, in
which each input parameter is perturbed along as discretized input space while
fixing others. In the MOAT, the $k$-dimensional input space (for $k$ parameters)
is partitioned uniformly in $p$ levels, creating a grid with $p^k$ points in
which the model evaluations take place. Each perturbation of an input parameter 
$x_i$ creates a parameter elementary effect computed as
shown in Equation~\ref{eq:ee}.

\begin{equation}
EE_i = \frac{y(x_1, ..., x_i+ \triangle_i, ...,x_k)-y(x)}{\triangle_i}
\label{eq:ee}
\end{equation}

where y(x) is the image analysis output metric value before the
perturbation. In our case, the output refers to the metric of interest
calculated comparing the application output mask to a reference mask. The
reference mask set, in these analyses, was calculated using the application
default parameters. Therefore, we will evaluate how the changes in the input
parameters reflect in mask changes that are computed using different spatial
metrics of interest that compare masks using a fixed pre-computed mask set as a
reference.  To account for global SA, the $\triangle_i$ value is typically
large.  This implementation uses $\triangle_i = \frac{p}{2(p-1)}$ that leads to
steps slightly larger that half of the input range for input variables scaled
between 0 and 1. 

The distribution of $r$ elementary effects of the input space characterizes the
effects on the output, which are measured using mean ($\mu$), modified mean
($\mu^*$), and standard deviation ($\sigma$) of elementary effects for each
input parameter~\cite{Campolongo20071509}.  The mean and modified mean
represent the effects of the input on the output, whereas the standard
deviation reveals nonlinear effects. This analysis involve $n = r(k+1)$
application runs (or evaluations in the input parameter space), and $r$ value
is suggested to be on the range of 5 to
15~\cite{Bertrand-2015,Alam:2004:UMR:1161734.1161926}. The MOAT does not
produce information about interactions of parameters, but gives evidence of
their existence.

\subsubsection{Methods to Compute Importance Measures}
These methods provide quantitative measurements of the correlations between
input parameters and application output or different input parameters through
correlation coefficients. If a sample of application runs for a set of input
values and the respective output measures are available, a linear model could
be fit on the sample to try explaining it. Some of the coefficients that we
compute out of these analyses include Pearson's correlation coefficient (CC)
and partial (PCC) correlation coefficient as well as Spearman's rank
correlation coefficient (RCC) and partial rank correlation coefficient (PRCC) .
The simple and partial correlation coefficients are similar, but the latter
excludes effects from other parameters or variables.  When the variables are orthogonal, the
simple and partial correlations are the same, whereas the ranked correlations
are helpful when the relationship between parameters are non-linear~\cite{opac-b1096602}.

The CC for  $x$ and $y$ is calculated as: $Corr(x, y) = r_{xy} =
\frac{\sum\nolimits_{i}(x_i -
\bar{x})(y_i-\bar{y})}{\sqrt{\sum\nolimits_{i}(x_i -
\bar{x})^2\sum\nolimits_i(y_i-\bar{y})^2}}$, where $x$ and $y$ could be two
parameters, a parameter and the output, etc. The output here is calculated
similarly to that described in the previous section. A pre-computed mask set is
used as a reference, and the resulting mask set for a set of input parameters
is compared to the reference set and the output of the comparative analysis is
taken as the result of the application of that particular parameter set. The
Spearman's correlation is similar to the Pearson's, except that it is
calculated based on ranked data~\cite{saltelli2004sensitivity}. The application
runs or sample used to calculate these metrics is typically generated from a
probabilistic exploration of the parameter space. We support a number of
stochastic methods to perform the exploration, such as the commonly used Monte
Carlo (random) sampling and Latin hypercube sampling(LHS). The LHS have been
shown to achieve better accuracy in parameter sensitivity
studies~\cite{10.2307/1268522}.  

Our implementation also supports the Variance-based Decomposition (VBD) global
sensitivity method. VBD splits output uncertainty effects among individual or
groups of parameters~\cite{Weirs2012157} and can account for non-linear
relationships among them.  VBD computes the ``main effect'' sensitivity index
$S_i$~\cite{Sobol2001271} and the ``total effects'' sensitivity index
$S_{T_i}$~\cite{Saltelli2002280}. The $S_i$ measures the amount of the output
variance that can be attributed to parameter $i$ alone (first-order effects).
If the sum of the $S_i$ values is close to one, most of the output variance is
explained by single parameter effects. The total effect index $S_{T_i}$
measures the first-order and higher-order effect due to $i$ interaction with
other parameters. The possible number of higher-order effects grows
exponentially with the number of input parameters. Therefore, in practice, it
is not viable to account for interactions of order higher than two. The VBD is
very compute intensive with respect to sampling, since for $k$ input parameters
and $n$ samples, it requires $n(k+2)$ application runs, and reasonably accurate
indices require $n$ to be in the other of hundreds or thousands. Those indices
are computed here as described in~\cite{Weirs2012157}.

\subsection{Parameter Auto-tuning Algorithms} \label{sec:auto-tuning}
%
%
%
The auto-tuning execution flow (presented in Figure~\ref{fig:overview}) is
similar to that of parameter study. The auto-tuning algorithm selects one or
more sets of application parameter values. The image analysis application is
executed for those parameter sets to compute the metric of interest against a
reference dataset (e.g., an image segmentation mask annotated by a
pathologist). The metric output for each parameter set is fed back to the
auto-tuning algorithm, which computes another set(s) of parameters to be
evaluated.  This iterative process continues until one of the two supported stop
conditions is met: (i)~maximum number of iterations or (ii)~a given metric
threshold value is reached.  

%

The current implementation supports the following auto-tuning algorithms:
Nelder-Mead simplex (NM)~\cite{6012898}, Parallel Rank Order
(PRO)~\cite{6012898}, and a Genetic Algorithm (GA)~\cite{sareni1998fitness}.
The NM and PRO approaches have been implemented using Active Harmony
(AH)~\cite{6012898,Tabatabaee:2005:PPT:1105760.1105822}, whereas the GA has
been developed by us. Active Harmony is an auto-tuning system that is primarily
designed and employed for tuning application runtime performance. In our work,
we make a novel use of Active Harmony for searching parameter space to improve
application analysis results. All algorithms try to minimize (or maximize by
negation) an unknown function by probing and exploring the parameter search
space.
 
\emph{The Nelder-Mead} method uses a simplex or polytope of $k+1$ vertices in a
k-dimensional search space. The simplex is updated in each iteration by
removing the vertex with the worst value ($v_r$), which is replaced with a new
vertex which has a lower function value. This operation involves computing the
centroid $c$ of the remaining simplex vertices to replace $v_r$ with a point on
line $v_r + \alpha(c - v_r)$. Typical $\alpha$ values are 0.5, 2, and 3. The
values of $\alpha$ define whether the transformation on the simplex is a
reflection ($\alpha=2$), an expansion ($\alpha=3$), or a contraction
($\alpha=0.5$).  The Nelder-Mead method usually performs a reflection first
and, depending on the results, follows the reflection with a expansion or
contraction. The original method has been modified in Active Harmony to deal
with non-continuous search spaces.


\emph{The Parallel Rank Order (PRO)} algorithm uses a set of K points from a
simplex (K $\ge$ N + 1) for a N-dimensional space. Each iteration of the
algorithm calculates up to $K - 1$ new vertices, which are computed by a
reflection, expansion, and shrinking of the simplex around its vertex with the
optimal value. Multiple vertices generated during each iteration may be
evaluated independently. This feature enables concurrent execution of multiple 
copies of the application with different parameter sets (vertices) on a 
high performance machine. The reflection step succeeds if at least one of the evaluated
vertices lead to an improvement of the optimization results. If no point
succeeds during reflection, the simplex shrinks around the best vertex.  The
expansion check follows a successful reflection and is executed to accept the
simplex or not. The simplex is expanded when accepted.  Search continues with a
new iteration. The algorithm stops when it converges to a point in the search
space or after a number of iterations have been executed. 

\emph{The Genetic Algorithm (GA)}~\cite{sareni1998fitness} models points in the
search space (parameter values) such that each parameter corresponds to a gene
of an individual. The initial population in our implementation is created by
randomly selecting parameter values from the search space. The population is
evolved through reproduction (selection), crossover and mutation. The selection
phase retrieves a percentage of the individuals with the best fitness (i.e.,
the best results with respect to our optimization function) that are duplicated
to replace the ones with the worst fitness. In the crossover phase, the
individuals are grouped into pairs, and all genes with an index higher than a
certain value are swapped between the individuals. This index value is randomly
selected for each pair of individuals. Finally, each individual may suffer
mutation of its genes. In this case the new value for a gene is randomly
selected. After these phases, the new population is evaluated via a fitness
function~\cite{sareni1998fitness} (metric of interest).  The outcome of the
evaluation is input back to the algorithm to build another generation. The
process continues until a preset number of generations (iterations) is reached.
As a performance optimization, multiple individuals from a generation can be
evaluated concurrently.
%

Most prior works on the problem of parameter estimation or
optimization in imaging segmentation employ techniques
for specific segmentation models. A pseudo-likelihood is used in~\cite{1238478}
to estimate parameters for a conditional random field based algorithm. Graph
cuts are employed to compute maximum margin efficient learning in segmentation
parameters~\cite{Szummer:2008:LCU:1479250.1479296}. A non-convex function is
optimized in~\cite{miccai2007c,McIntosh2009} with techniques to avoid
sensitivity in segmentation due to the initial parameter choices.
Open-Box~\cite{journals/tvcg/SchultzK13} is another interesting solution that
is specific to segmentation algorithms based on spectral clustering. It deals
with the optimization by exposing key components of the segmentation to the
user. 
The Tuner system~\cite{Torsney-Weir:2011:TPP:2068462.2068683} deals with
general segmentation algorithms and focuses on creating statistical models from
sampling runs that describe the segmentation response function, and employs a
Gaussian process model to explore the parameter space.  The areas of the search
space identified as promising are further explored to refine the parameters. In
our work, we treat the segmentation algorithm as a black-box.  Additionally, 
instead of using a
statistical model that describes the application, we employ efficient
optimization algorithms that can quickly converge to desired results.
Our solution also allows for the use of multiple auto-tuning algorithms. 
We employ HPC with several optimizations to accelerate application runs in 
the parameter tuning phase. We provide to the
user several domain specific metrics for evaluating algorithm outputs, and
new metrics can be incorporated via a spatial querying engine.

\subsection{Scalable Execution of Sensitivity Analysis and Auto-tuning on Parallel Machines}
\label{sec:rt-framework}
%
%
Sensitivity analysis and parameter auto-tuning are extremely time consuming processes, because 
computationally expensive analysis pipelines have to be executed multiple times on large volumes 
of imaging data. They can benefit from high performance computing (HPC) systems that have hybrid 
computation nodes equipped with accelerators (GPUs and Intel Xeon Phi co-processors) and multiple 
levels of data storage consisting of GPU and CPU memories, SSDs, local spinning disks.  
The use of hybrid systems equipped with CPU and accelerators is a fast growing
topic that has attracted attention for research in are as compiler techniques,
scheduling, runtime systems, and parallelization of
applications~\cite{1504194,6061070,ravi2010compiler,hpdc10george,6267858,Teodoro:2012:Cluster,augonnet:hal-00725477,Rossbach:2011:POS:2043556.2043579,opencv_library,cluster09george,Teodoro:2014:CPA:2650283.2650645,teodoro2015application}.
In special, several Biomedical Informatics applications and research
initiatives are able to benefit from accelerators and parallel HPC
systems~\cite{Foran403,hpdc10george,Liocv146,Li363,Lindberg197,Kaspar535,4814676,Eliceiri,Fang2013184,10.1371,citeulike:7266282,Teodoro:2013:Parco,Teodoro-2012-Morph,Teodoro:2008:IJPP,10.1109/CCGRID.2007.20}.
In this work, we have implemented support for sensitivity analysis and parameter auto-tuning on the region templates (RT) runtime 
system~\cite{Teodoro2014589} to address the computational requirements of these processes on HPC systems with co-processors.  
The region templates runtime system provides core functions for scheduling of application operations 
across multiple computation nodes and on co-processors (such as CPUs and Intel Xeon Phis) and manages 
data movement across multiple storage layers.   

The processing
structure of a region template application is expressed as a hierarchical
dataflow graph. That is, an operation itself can be composed of lower-level
operations organized into another dataflow. Application workflows are decomposed
into components of computation that consume, transform, and produce region
template data objects instead of reading/writing binary data directly from/to
other tasks or disk. The region template data abstraction provides a generic 
container
template for common data structures, such as pixels, points, arrays (e.g., images or
3D volumes), segmented and annotated objects and regions, that are defined in a
spatial and temporal domain. Using these containers, the application developer can 
avoid having to manually deal with communication of data structures accross memory 
hierachies of nodes in a distributed machine.

\begin{figure}[htb!]
\begin{center}
\includegraphics[width=0.77\textwidth]{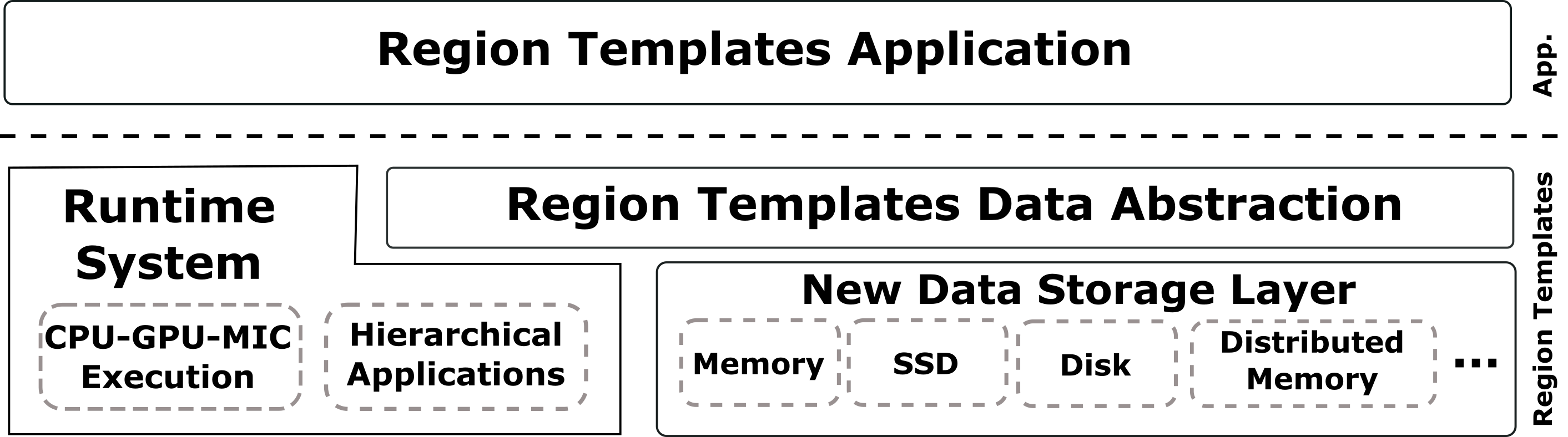}
\vspace*{-2ex}
\caption{Architecture of the framework.}
\label{fig:rt-arch}
\end{center}
\end{figure}

Figure~\ref{fig:rt-arch} presents the main components of the region template
framework: the region templates data abstraction, the runtime environment, and the
hierarchical data storage layer. The runtime environment deals with the
instantiation of components for execution with nodes of a distributed memory
environment and interacts with the data storage layer for data region
management. The storage layer is responsible for exploiting the memory
hierarchy in a distributed memory system to efficiently move region template
data among nodes.  The runtime environment of RT implements a Manager-Worker 
execution model that combines a bag-of-tasks execution with the dataflow pattern.
The application Manager creates instances
of  coarse-grain stages, which include \emph{input data regions}, and exports
the dependencies between the stage instances.  The
assignment of work from the Manager to Worker nodes is performed at the
granularity of a stage instance using a demand-driven mechanism.
Each Worker
may execute several stage instances concurrently to take advantage of multiple
computing devices (CPUs and co-processors) on a computation node. Computing 
devices on a node are used cooperatively by dispatching fine-grain tasks for 
execution on a CPU core or a co-processor (e.g., an Intel Phi or a GPU) via a 
performance-aware task scheduling (PATS) algorithm~\cite{Teodoro-IPDPS2013,Teodoro-IPDPS2012}. 
PATS assigns tasks to a CPU core or an accelerator based on the tasks estimated 
acceleration on each device and on the device load. We refer the reader to our 
earlier publication~\cite{Teodoro2014589} for details of the region templates 
framework and runtime environment. In this paper we describe two optimizations 
in the region templates framework that target the requirements of 
the sensitivity study and parameter tuning processes. The first one is a new hierarchical
data storage layer, which improves data
locality for this class of application (Section~\ref{sec:caching}). The second optimization is a
strategy to avoid re-computation overheads during the evaluation
of multiple parameter sets for the same workflow
(Section~\ref{sec:auto-parallel}). 

\subsubsection{Storage Management and Optimizations} 
\label{sec:caching}
The same or overlapping sets of data elements in one
or more stages of the analysis workflow are processed and re-processed 
as input parameters are varied in an algorithm sensitivity analysis 
run. To
take advantage of this, we have developed a new hierarchical
storage infrastructure for RT and a strategy that considers data locality
during the scheduling of application stage instances to improve the effectiveness
of maintaining data in faster memory for reuse.
%
%
The data storage layer is built as a distributed data management
infrastructure, which can use an arbitrary number of memory layers within a
node and across a distributed memory system. It is implemented as a module of
the Worker processes in the runtime system. The memory/storage hierarchy of the
target system is defined in a configuration file. The configuration file
includes the number of storage levels used, the position of each storage
in the hierarchy and a description of each level: type of storage
device (e.g., RAM, SSD, HDD, etc), storage capacity, path for storing data, and
storage visibility (local or global). Storage specified as {\em local} can only
be directly accessed within the node (Worker process); Data regions associated
with local storage are not directly visible to other nodes (Workers). Storage
specified as {\em global} is visible to other nodes and is
used to exchange data among stages of a region templates
application. 

The data storage layer is in charge of storing and retrieving instances of 
region templates.  The runtime system contacts the data storage layer
whenever a region template instance is output or requested by an application 
stage or operation. When accessing a region template instance, the search for the requested 
data may result in three cases: (i)~The data is found in
a local storage component; it is directly returned to the application stage/operation. 
(ii)~The data is
found in global storage; it is retrieved and transferred by the storage layer to 
the requesting node and application stage/operation. (iii)~The data is 
found neither in local storage on the requesting node nor in global storage. 
This means the data is stored in local storage of the node in which
it was produced (the source node). Inter-processor communication is necessary with the source 
node to move the data to a storage component of global visibility, before 
the data can be retrieved.  

Data regions are staged to the data storage layer automatically. The runtime
system iterates through data regions generated and used by an application
operation and inserts those marked as output to the storage layer. The
insertion is always performed into the highest (i.e., the fastest) level of the
storage hierarchy with enough capacity to save the data regions.  When a level
reaches its maximum storage capacity, a cache replacement strategy is employed
to select data regions that should be moved to a lower level in the hierarchy.
Each level of a storage hierarchy may use one of the supported data replacement
policies: First-In, First-Out (FIFO) and Least Recently Used (LRU).  New
policies can be incorporated via the application programming interface.

%
%
Data should ideally be reused or retrieved when it resides in faster memory/storage layers. 
In order to reduce data access costs, we have developed a data locality-aware scheduling (DLAS) 
approach that considers the location of data to be accessed when scheduling and mapping 
application stages and operations to the nodes of the computation system. 

The DLAS strategy is implemented at the Manager level of the runtime system.
With this policy, when the notification is received that a given application
stage instance (referred to as the original stage instance) has finished, the
Manager takes into account the locality of the data produced by that stage
instance to determine the node in which stage instances that use the produced
data should be executed. In this process, DLAS calculates the amount of data
reuse of stage instances that consume data from the original stage instance,
and inserts them into a queue of preferred stage instances for execution in the
Worker node that executed the original stage instance. A queue of preferred
instances is maintained for each Worker in decreasing order of the amount of
expected data reuse. When a Worker requests a stage instance for execution, the
Manager will try to assign the stage instance that reuses the maximum amount of
data --- that is, the stage instance on the top of the queue for the requesting
Worker. If the queue is empty or none of the stage instances in the queue have
dependencies resolved (i.e., they cannot be scheduled for execution), an
instance is chosen using the First-Come, First-Served (FCFS) order among those
stage instances ready for execution.


%
%
\subsubsection{Optimized Simultaneous Parameter Evaluation} \label{sec:auto-parallel}
The execution of parameter study and auto-tuning algorithms (PRO and GA) allows
for multiple simultaneous parameter set executions and evaluations per
iteration of the process (Figure~\ref{fig:overview}). We exploit this
characteristic to optimize the evaluation of these multiple runs with different
parameter values. Instead of replicating the segmentation/application workflow
for each parameter set to be evaluated, we created a strategy that merges and
eliminates replicas of common computation paths of the workflows for faster
evaluation of multiple parameters in each iteration.  


\begin{figure}[htb!]
\begin{center}
        \includegraphics[width=0.77\textwidth]{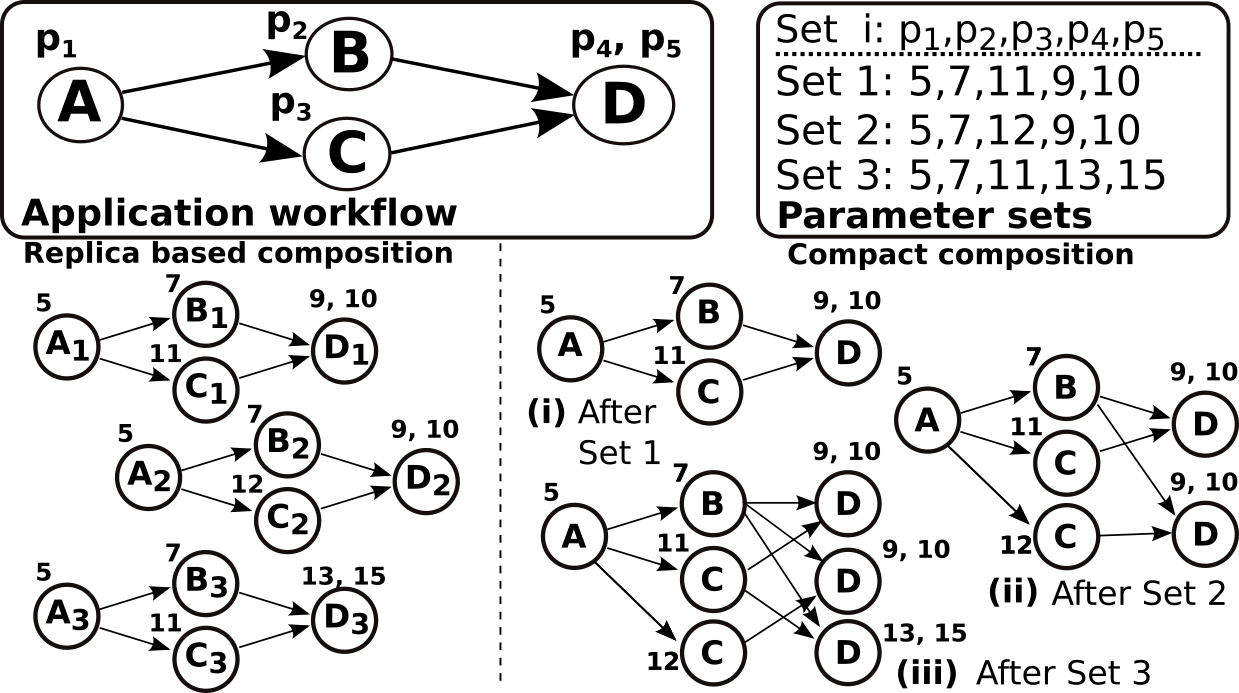}
\vspace*{-3ex}
\caption{Application composition schemes.}
\label{fig:app-graph}
\end{center}
\end{figure}

Figure~\ref{fig:app-graph} shows two schemes for instantiating an application
workflow in a parameter analysis study in which multiple parameter sets are
tested in an iteration. The {\em replica based scheme} instantiates the entire
application workflow for each parameter set and dispatches multiple independent
workflow instances for execution. The {\em compact composition scheme}, on the
other hand, merges the instances of an application workflow into a single,
compact workflow graph to reuse common stages in the separate workflow
instances.  The compact workflow graph representation draws from a data
structure called FP-Tree~\cite{Han:2000:MFP:342009.335372}. The FP-tree
represents sets of transactions in a structure in which common parts of
transactions are expressed in a single path on the structure. In our case, we
want to merge multiple workflows to create another workflow in which common
computations from multiple parameters are eliminated. The common computations
are found in stages that have the same parameters and input data in multiple
workflows. 

\begin{algorithm}
\small
  \caption{Compact Graph Construction
    	\label{alg:comp-graph}}
  	
	\begin{algorithmic}[1]
	\State {\bf Input:} appGraph; parSets;
	\State{\bf Output:} comGraph;
	\For{{\bf each} set $\in$ parSets}
		\State appGraphInst = \Call{instantiateAppGraph}{set};
		\State \Call{MergeGraph}{appGraphInst.root, comGraph.root};
	\EndFor
	\Procedure{MergeGraph}{appVer, comVer}
		\For{{\bf each} v $\in$ appVer.children}
			\If{(v' $\gets$ find(v, comVer.children))}
				\State \Call{MergeGraph}{v, v'};
			\Else
				\If{((v' $\gets$ PendingVer.find(v))=$\emptyset$)}
					\State v' $\gets$ clone(v)
					\State v'.depsSolved $\gets$ 1
					\State comVer.children.add(v')
					\If{v'.deps $\ge$ 1}
						\State PendingVer.insert(v')
					\EndIf
					\State \Call{MergeGraph}{v, v'};
				\Else
					\State comVer.children.add(v')
					\State v'.depsSolved $\gets$ v'.depsSolved+1
					\If{v'.depsSolved = v'.deps}
						\State PendingVer.remove(v')
					\EndIf
					\State \Call{MergeGraph}{v, v'}
				\EndIf
			\EndIf		
		\EndFor
	\EndProcedure
  \end{algorithmic}
\end{algorithm}
\begin{sloppypar}

The construction of the compact representation is presented in
Algorithm~\ref{alg:comp-graph}. To simplify the presentation of the algorithm,
we assume that both the compact graph (comGraph) and each instance of the
application graph (appGraphInst) to be merged have a single start root node.  
The algorithm takes the application workflow graph (appGraph) and parameter
sets to be tested simultaneously as input (parSets) and outputs the compact
graph (comGraph). It iterates over each parameter set (lines 3-5)
to instantiate a replica of the application workflow graph with parameters from
$set$. It then calls {\scshape MergeGraph} to merge the replica to the compact
representation.  \end{sloppypar}

The {\scshape MergeGraph} procedure walks simultaneously in an application workflow
graph instance and in the compact representation. If a path  in the application
workflow graph instance is not found in the latter, it is added to the compact graph.
The {\scshape MergeGraph} procedure receives the current set of vertices in the
application workflow (appVer) and in the compact graph (comVer) as a parameter
and, for each child vertex of the appVer, finds a corresponding vertex in the
children of comVer. Each vertex in the graph has a property called {\em deps}, which
refers to its number of dependencies. The find step considers the name of a
stage and the parameters used by the stage. If a vertex is found, the path already
exists, and the same procedure is called recursively to merge sub-graphs
starting with the matched vertices (lines 8-9). When a corresponding vertex is
not found in the compact graph, there are two cases to be considered (lines
10-23). In the first one, the searched node does not
exist in comGraph. The node is created and added to the compact graph (lines
11-17). To check if this is the case, the algorithm verifies if the node ($v$)
has not been already created and added to comGraph as a result of processing
another path of the application workflow that leads to $v$. This occurs for
nodes with multiple dependencies, e.g., D in Figure~\ref{fig:app-graph}. If the
path (A,B,D) is first merged to the compact graph, when C is processed, it
should not create another instance of D.  Instead, the existing one should be
added to the children list as the algorithm does in the second case (lines
19-23). The PendingVer data structure is used as a look-up table to store such
nodes with multiple dependencies during graph merging. 

In the proposed algorithm, application workflows are expected to be directed
graphs. If a stage is used multiple times in the same application, it
is necessary to rename the repeated stage to prevent the algorithm from
finding incorrect nodes in the look-up of pending nodes (line 11).

\subsubsection{Spatial Comparison Support} \label{sec:comparative-analysis}

We have implemented a comparative analysis module in the region templates system
for efficient quantitative comparison of multiple segmentation results using a
query-based interface. By performing comparison of objects from multiple image
segmentation results using a set of core operations that include spatial
cross-matching, overlay of objects, spatial proximity computations between
objects, and global spatial pattern discoveries, we are able to compute a
number of metrics of interest from objects detected in segmentation runs that
are used in algorithm sensitivity analysis. The metrics we built in region
templates using this module include: Dice Coefficient, Jaccard Coefficient,
Intersection Overlapping Area, Non-Overlapping Area (Number of pixels
differently segmented)~\cite{TA2015}. However, the core operations previously
described can be combined in a number of ways to create a more extensive set of
metrics. Also, we support more complex queries that include spatial proximity
computations between objects using Nearest Neighbor Queries (KNN), which may be
combined with any other metric of interest to investigate the characteristics of
a subset of objects close to another object of interest (e.g, cells close to a
vein or other structure). These features are available with region templates
through its integration with a GIS querying framework~\cite{aji2013hadoop,aji2014haggis}. In
order to use it in an application, the user simply connects an existing
application stage (component) developed in RT to her application, passing the
masks that need to be compared to this component.

\begin{figure}[htb!]
\begin{center}
        \includegraphics[width=\textwidth]{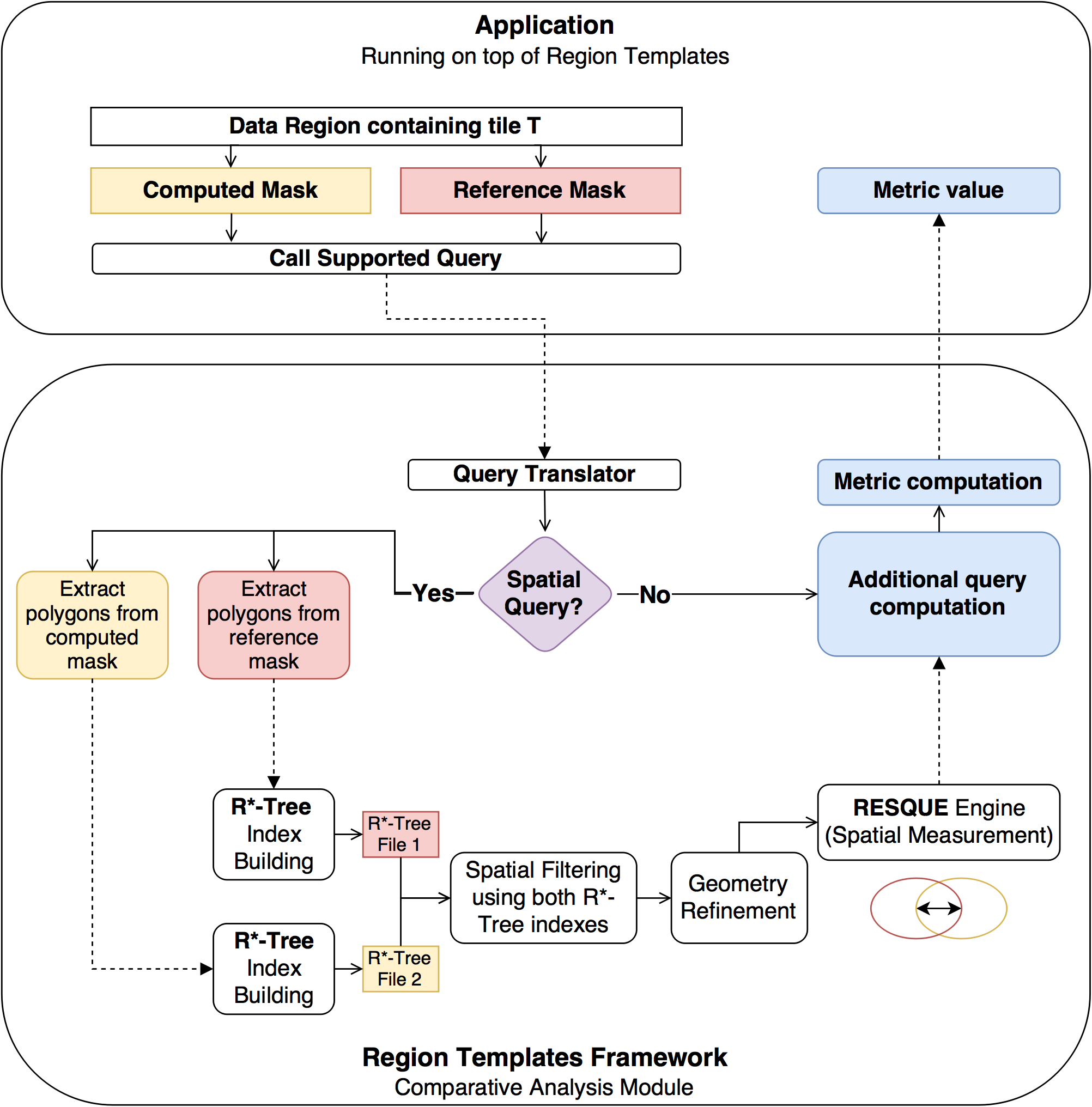}
\vspace*{-1ex}
\caption{Metric Computation Workflow.}
\vspace*{-3ex}
\label{fig:comparative-analysis}
\end{center}
\end{figure}

The overall view of the metric computation flow is presented in
Figure~\ref{fig:comparative-analysis}. The RT application computes a mask and
passes the computed mask and any other reference mask as input to the
comparative analysis module, which as described is instantiated in the
distributed environment an application stage. In order to execute spatial
queries, the Comparative Analysis module extracts the objects (cells, veins,
etc) from the masks and convert them into polygons.  After this process, an
application running under Region Templates Framework can use one of the
supported queries to perform spatial processing. The implemented queries can
use the core capabilities of the  Hadoop-GIS Framework~\cite{aji2013hadoop} to
calculate the intersection area between the two sets or calculate their spatial
proximity. After receiving the results from the spatial processing engine, the
Comparative Analyses module may calculate the Dice Coefficient, Jaccard Index
or some other metric in order to complete the query task.

The actual spatial queries are computed using provided  Hilbert R*-Tree
indexing-driven spatial query engine, which is critical to quickly identify
overlapping objects in join based operations~\cite{beckmann1990r}. First, a
R*-Tree is built from objects minimum bounding boxes in each mask, and a
spatial filtering operation is performed to identify possibly overlapping
objects (those with intersecting bounding boxes), which are refined to those
that overlap. This set is passed to a final phase in which actual operations
performed to compute spatial measurements results. These results may be the
overlapping area of the intersection, the area of the union of the polygons etc. A detailed explanation of the coefficients we
calculate is provide bellow:

\paragraph {Dice Coefficient or The S{\o}rensen-Dice index
\cite{sorensen1948method, dice1945measures}}  This coefficient ranges from 0 to
1 and is used to measure the similarity of two samples. This metric can be
calculated by dividing twice the intersection area of two samples by the sum of
their respective areas. 

\begin{figure}
\begin{center}
        \includegraphics[width=\textwidth]{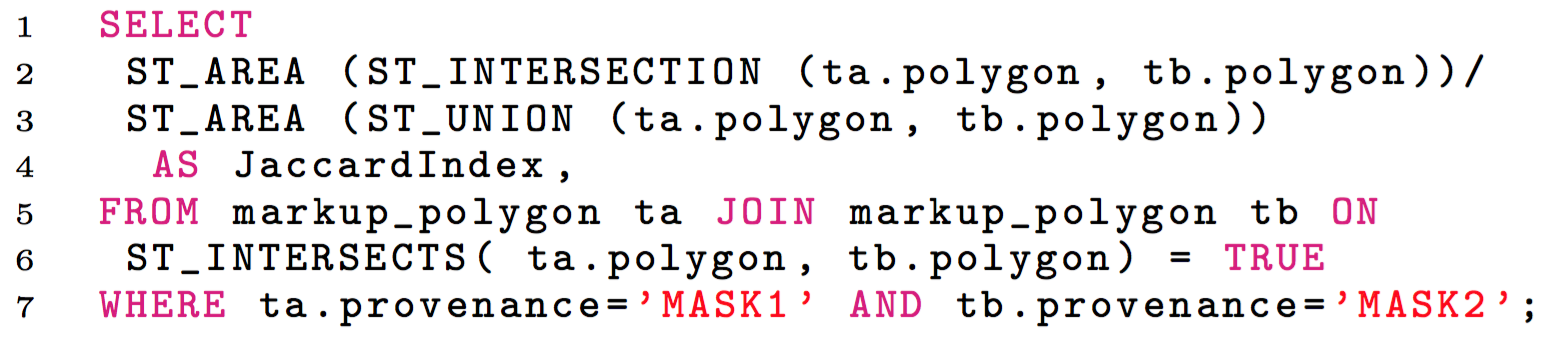}
\vspace*{-1ex}
\caption{An equivalent SQL query to the Jaccard Index Metric.}
\vspace*{-3ex}
\label{fig:sql-query}
\end{center}
\end{figure}

\paragraph {Jaccard Index Metric} it also ranges from 0 to 1 and is similar to
the Dice Coefficient. The Jaccard Index is calculated by dividing the
intersection area between two sets by the union area of these
sets~\cite{jaccard1901etude}. The SQL expression equivalent to this metric is
shown in Figure~\ref{fig:sql-query}.

\paragraph{Intersection Overlap Area Metric} This metric represents the
relation between the total area of the intersection between two masks divided
by the area of reference mask set.

\paragraph{Nearest Neighbor Query} Nearest neighbor (KNN) search in analytical
imaging analysis can be computationally expensive, we use Hadoop-GIS to
efficiently perform these queries in order to understand correlations between
spatial proximity and cell features. The implemented query allows selecting the
K-nearest objects or within a certain bound. A typical scenario that a
application can take advantage of this query is to find the nearest objects
(blood vessels, cells, etc) for each cell of a given set in order to study
their characteristics and correlations.

\section{Results and Discussion} \label{sec:results}
We have evaluated the proposed methods and optimizations using a set of Glioblastoma
brain tumor tissue images collected in brain cancer
studies~\cite{kong2013machine}. The images were divided into 4K$\times$4K
tiles; each tile was processed concurrently with other tiles in a bag-of-tasks
style execution.  The image analysis workflow consisted of normalization,
segmentation and comparison stages. The comparison stage computes the
difference between masks generated by the application and the reference data
for each set of parameters.  The experimental evaluations were conducted on two
distributed memory machines.  The first one is the TACC Stampede
cluster
has a dual socket Intel Xeon E5-2680 processors, an Intel Xeon Phi SE10P
co-processor and 32GB RAM. The nodes are inter-connected via Mellanox FDR
Infiniband switches.  The second cluster, called Eagle, is a small cluster at
Stony Brook University with 10 computing nodes. Each node has dual socket Intel
Xeon E5-2660 processors, an Intel Phi 5110P co-processor, 256GB RAM, 1TB
spinning disk and a 512GB SSD. Stampede uses a Lustre file system accessible
from any node, while each node on Eagle has a local Linux file system.  The
application and middleware codes were compiled using Intel Compiler 13.1 with
``-O3'' flag.  The MIC operations used the {\em offload mode} -- a computing
core was reserved to run the offload daemon and at most 240 threads were
launched. 

\subsection{Sensitivity Analysis}

\subsubsection{Finding Important Parameters with MOAT}

The use-case microscopy image analysis workflows employed in this study have
k=15 and k=7 parameters (described in Table~\ref{tab:parameters}), which we
first study using MOAT in order to try pruning the parameter space before more
costly analyses such as VBD and tuning are employed. Thus, in this phase, we
use the MOAT design to select the most impacting parameters to be used in
downstream analyses. The output of the applications to the MOAT method is the
difference in number of pixels from the mask computed with parameter values
selected by the method and a mask precomputed using the application default
parameter values. The watershed based workflow employs 55 Glioblastoma brain
tumor~WSIs (4,276 4K$\times$4K non-background only image tiles) and the level
set based is evaluated with 1 Glioblastoma brain tumor~WSI (71 4K$\times$4K
that were partitioned into 512$\times$512 tiles). A smaller dataset is used
with the level set workflow because it is more compute demanding.

We have considered parameter space partition with 20 levels for each of the $k$
parameters in Table~\ref{tab:parameters}. The number of executions of the image
analysis workflow for the entire input dataset value is calculated as
$n=r(k+1)$. The experiments were performed using 128 nodes of the Stampede
cluster and they 15681s and 6825s, respectively for the watershed and level set
workflows when r is 15.

\definecolor{maroon}{rgb}{0.8, 0.0, 0.0}
\definecolor{green}{rgb}{0.0, 0.5, 0.0}
\definecolor{yellow}{rgb}{1.0, 0.88, 0.21}

\begin{table}[h!]
\begin{center}
\vspace*{-2ex}
\caption{MOAT analysis for both segmentation workflows and r values of 5, 10, and 15. We classify
in green, yellow, and red, respectively, those parameters have high, medium,
and low effect on the output. Several parameters have presented non-linear
effects in the analysis (high $\sigma$ value), and as such we conservatively
pipeline input parameters with high and medium effects to further and more
costly studies.}
\begin{scriptsize}
\subtable[MOAT results for the watershed based segmentation workflow.]{\label{tab:moat-water}
\begin{tabular}{c c c c c c c}
\hline

\multirow{2}{*}{Parameter}	& \multicolumn{2}{c}{r = 5} 	& \multicolumn{2}{c}{r=10} & \multicolumn{2}{c}{r=15}		\\ \cline{2-7}
				& $\mu^*$	& $\sigma$	& $\mu^*$	& $\sigma$	& $\mu^*$	& $\sigma$	\\ \hline
\rowcolor{maroon!30}1 (Blue)	&4.59E+04	&1.03E+05	&2.30E+04	&7.26E+04	&2.77E+04	&7.87E+04	\\ \hline
\rowcolor{maroon!30}2 (Green)	&3.50E+04	&7.83E+04	&3.89E+04	&9.20E+04	&2.59E+04	&7.38E+04	\\ \hline
\rowcolor{maroon!30}3 (Red)	&0.00E+00	&0.00E+00	&0.00E+00	&0.00E+00	&0.00E+00	&0.00E+00	\\ \hline
\rowcolor{maroon!30}4 (T1)	&5.90E+05	&9.01E+05	&1.46E+07	&4.24E+07	&2.01E+07	&4.15E+07	\\ \hline
\rowcolor{yellow!30}5 (T2)	&7.10E+07	&1.24E+08	&4.63E+07	&9.42E+07	&5.46E+07	&9.08E+07	\\ \hline
\rowcolor{green!30}6 (G1)	&3.78E+08	&5.87E+08	&1.10E+09	&2.25E+09	&9.35E+08	&1.93E+09	\\ \hline
\rowcolor{green!30}7 (G2)	&3.08E+09	&3.67E+09	&2.38E+09	&2.97E+09	&2.78E+09	&3.83E+09	\\ \hline
\rowcolor{green!30}8 (MinSize)	&2.26E+08	&3.24E+08	&8.79E+08	&2.44E+09	&6.52E+08	&2.00E+09	\\ \hline
\rowcolor{yellow!30}9 (MaxSize)			&7.32E+07	&1.20E+08	&8.18E+07	&1.15E+08	&7.30E+07	&1.07E+08	\\ \hline
\rowcolor{yellow!30}10 (MinSizePl)			&8.82E+07	&1.32E+08	&9.35E+07	&1.30E+08	&1.27E+08	&1.74E+08	\\ \hline
\rowcolor{yellow!30}11 (MinSizeSeg)			&1.64E+08	&2.84E+08	&1.92E+08	&3.05E+08	&1.61E+08	&2.58E+08	\\ \hline
\rowcolor{maroon!30}12 (MaxSizeSeg)&5.46E+06	&9.33E+06	&6.11E+06	&9.72E+06	&2.00E+07	&6.04E+07	\\ \hline
\rowcolor{maroon!30}13 (FillHoles)&5.13E+06	&9.44E+06	&6.95E+06	&1.18E+07	&7.15E+06	&1.18E+07	\\ \hline
\rowcolor{green!30}14 (Recon)	&1.62E+09	&3.12E+09	&3.15E+09	&4.76E+09	&2.89E+09	&4.39E+09	\\ \hline
\rowcolor{maroon!30}15 (Watershed)&5.06E+07	&5.88E+07	&6.05E+07	&6.44E+07	&5.37E+07	&5.36E+07	\\ \hline
\end{tabular}
}
\subtable[MOAT results for the level set based segmentation workflow.]{\label{tab:moat-level}

\begin{tabular}{c c c c c c c}
\hline

\multirow{2}{*}{Parameter}	& \multicolumn{2}{c}{r = 5} 	& \multicolumn{2}{c}{r=10} & \multicolumn{2}{c}{r=15}		\\ \cline{2-7}
				& $\mu^*$	& $\sigma$	& $\mu^*$	& $\sigma$	& $\mu^*$	& $\sigma$	\\ \hline
\rowcolor{green!30}1 (OTSU)	&9.64e+07	&1.35e+08	&1.12e+08	&1.19e+08	&1.05e+08	&1.19e+08	\\ \hline
\rowcolor{green!30}2 (CW)	&1.97e+07	&3.33e+07	&1.47e+07	&2.50e+07	&1.62e+07	&2.70e+07	\\ \hline
\rowcolor{yellow!30}3 (MinSize)	&5.89e+06	&7.42e+06	&6.37e+06	&7.62e+06	&6.06e+06	&7.32e+06	\\ \hline
\rowcolor{yellow!30}4 (MaxSize)	&1.77e+05	&3.94e+05	&5.24e+05	&1.06e+06	&1.12e+06	&3.15e+06	\\ \hline
\rowcolor{green!30}5 (MsKernel)	&2.96e+07	&4.23e+07	&3.34e+07	&4.43e+07	&3.17e+07	&4.15e+07	\\ \hline
\rowcolor{green!30}6 (LevelSetIt)&1.27e+07	&2.42e+07	&7.22e+06	&1.70e+07	&7.99e+06	&1.67e+07	\\ \hline
\rowcolor{maroon!30}7 (Dummy)	&1.93e+05	&3.48e+05	&1.96e+05	&3.84e+05	&1.81e+05	&3.58e+05	\\ \hline
\end{tabular}
}
\end{scriptsize}
\label{tab:moat}
\vspace*{-5ex}
\end{center}
\end{table}

The modified mean and standard deviation of the number of pixel difference are
presented in Table~\ref{tab:moat}. Although there is a variability in the
$\mu^*$ and $\sigma$ as $r$ increases, it does not significantly affect the
separation among parameters that are or not important. For the watershed based
workflow (Table~\ref{tab:moat-water}), it is possible to notice that parameters 6, 7, 8, and 14 are the
most important due their higher 
$\mu^*$ and $\sigma$ values. It is also interesting to note that most of the
input variables have non-linear interactions, because of the large value of
$\sigma$. For $G1$ and $MinSize$, it is the most important component.  This
indeed makes sense because $G1$ interacts with $G2$, as they are used to
determine a range threshold value.  Also, $MinSize$ interacts with other
parameters used to filter out objects that are not within a given range size
value. 

Most input variables have considerable interactions or non-linear effects,
which limits our ability to understand the actual effect each input to the
results.  Therefore, we have decided to make a conservative pruning of
parameters during this step, and we forward other parameters considered with
medium importance to more detailed and costly analysis. These parameters are 5,
9, 10, and 11, which have at least one component ($\mu^*$ or $\sigma$) higher
than $10^8$. The other ones have low linear or non-linear effects are discarded
at this stage.

The MOAT study of the level set based segmentation have included a dummy
parameter. This parameter is inputted in the MOAT but is not passed to the
application, and it is used here to quantify output segmentation differences
due to the algorithm stochastic nature. The de-clumping phase of the
segmentation is implemented using a randomized clustering strategy and, as a
result, segmentation outputs from two runs with the same parameters may differ.
As shown in Table~\ref{tab:moat-level}, although the OTSU ratio has the higher
effect in the output, all parameters seem to have significant effects with a
high non-linear interaction component. The only exception is the dummy
parameter, which has a low impact. It shows that the application output
variabilities due to its stochastic nature are minor as compared to the effects
of the parameters. For this workflow, we only prune the dummy parameter.

%


\subsubsection{Importance Measures}
%

\paragraph{Pearson's and Spearman's Correlation Coefficients} The first set of
experiments in this section computed the CC, PCC, RCC, and PRCC metrics.  It
was executed using LHS sampling with 400 points/samples, meaning the sample
segmentation workflow and comparison was executed 400 times for the same
datasets use in the MOAT analysis. The execution took 27078s and 22696s,
respectively, for the watershed and level set workflows using 128 computing
nodes.

\begin{table}[h!]
\begin{center}
\vspace*{-2ex}
\caption{Pearson's and Spearman's Simple and Partial Correlation Coefficients
of the segmentation workflows with respect to changes input parameters and
their impact to the output changes. The output changes are measured in terms
of the number of pixels modified in the segmentation output as parameter values
are changes.}

\begin{scriptsize}
\subtable[Results for the watershed based segmentation workflow.]{\label{tab:imp-water}
\begin{tabular}{c c c c c }
\hline
Parameter	&CC		&	PCC	& RCC		&	PRCC 	\\ \hline \hline
	  T2	&-4.94e-02	&-9.27e-02	& 3.97e-02	& 2.13e-02	\\ \hline
          G1 	& 1.01e-01 	& 1.57e-01   	& 1.41e-01    	& 1.83e-01	\\ \hline
          G2 	& 4.83e-01	& 5.08e-01  	& 3.74e-01    	& 3.99e-01	\\ \hline
     MinSize 	& 1.02e-01	& 1.25e-01  	& 1.38e-01    	& 1.36e-01	\\ \hline
     MaxSize 	& 7.72e-03	& 4.46e-02  	&-3.88e-02  	&-8.21e-03	\\ \hline
   MinSizePl 	& 8.79e-02	& 1.91e-02  	& 1.29e-01    	& 7.27e-02	\\ \hline
  MinSizeSeg 	&-6.29e-02	&-3.31e-02  	& 3.96e-02    	& 7.10e-02	\\ \hline
       Recon 	& 9.16e-02 	& 1.14e-01   	& 8.35e-02    	& 9.40e-02	\\ \hline
\end{tabular}}
\subtable[Results for the level set based segmentation workflow.]{\label{tab:imp-level}
\begin{tabular}{c c c c c }
\hline
Parameter	&CC		&PCC		& RCC		& PRCC 		\\ \hline \hline
OTSU	  	& 7.47e-01	& 7.52e-01	& 5.90e-01  	& 6.09e-01	\\ \hline
CW	   	&-5.18e-02	&-1.05e-01  	&-1.86e-01   	&-2.63e-01	\\ \hline
MinSize	   	& 5.05e-03	&-1.88e-02  	& 6.97e-02  	& 9.25e-02	\\ \hline
MaxSize		&-3.78e-02	& 5.93e-03  	&-9.35e-03   	& 2.84e-02	\\ \hline
MsKernel	&-3.74e-02	& 9.66e-02  	&-3.65e-02   	& 4.40e-02	\\ \hline
LevelSetIt 	&-6.50e-02	&-7.48e-02  	&-7.73e-02   	&-8.01e-02	\\ \hline
\end{tabular}}
\end{scriptsize}
\label{tab:ccs}
\end{center}
\end{table}

The correlations between parameters and output are
presented in Table~\ref{tab:imp-water} for the watershed workflow. We also compute simple correlations between
pairs of input parameters, but the data is omitted because of space
limitations. The analysis of the CC shows that most of the parameters have a
small (about 0.1) correlations with the output, with exception of the G2 whose
CC is 0.48. The differences from the CC and PCC values are an evidence of inter
parameter correlation effects.
The simple ranked based correlation (RCC) in four parameter is higher that the
Pearson's CC, which indicates that those parameters have a monotonic but not
linear correlation and may explain why a number of the parameters assumed small
CC values. Most of parameters hold a similar ranking (for instance using RCC)
as that shown in the MOAT. These analyses confirm most of the results attained
with MOAT, and are not useful in terms of doing an extra parameter pruning
before VBD is executed. However, the RCC shows a reduction on the gap
between most impacting other parameters, as compared to MOAT.

The correlations for the level set segmentation workflow are presented in
Table~\ref{tab:imp-level}, and it shows a number of interesting aspects. The
OTSU is highlighted as the most important parameter, and its CC and PCC values
are almost the same, indicating it effects are orthogonal with other
parameters. Although only OTSU appears to be important from CC, the PCC shows
that after excluding effects from other parameters the CW effect increases
significantly. The ranked correlation values (RCC and PRCC) also led to higher
values for OTSU and CW, and the same trend was observed between the simple to
partial correlations. The MaxSize parameter had low correlation values
and was excluded from the further analysis.

\paragraph{Variance-Based Decomposition (VBD)} The Sobol's indices are
presented in Table~\ref{tab:vbd} for the $k=8$ and $k=5$ input parameters,
respectively, for watershed and level set workflows resulting from previous
experiments. We have used the Satteli's formula~\cite{Saltelli2002280} with
Monte Carlo sampling, and each experiment required $N=n(k+2)$ application runs.
Because of the high computation costs, we have limited the value of $n$ to 200,
which seems to be sufficient because of the small variations observed when n
increases from 100 to 200.  The experiments with $n=200$ with the watershed
workflow requires 2,000 application runs for the 55~WSIs and took 150,890
seconds using 128 computing nodes with 820~TB of data read/staged during the
execution. The sequential execution of this experiment would take about
18~years. The execution with the level set took 211,912 seconds in 128
computing cores, which would take 25~years on a sequential execution.

The results in Table~\ref{tab:vbd-water} show that G2 is substantially more
impacting to the application output uncertainty than other input parameters.
However, a large fraction of the application output variance can not be
attributed to single input parameter effects, because the sum of the $S_i$
indices is considerably smaller than one (0.74 for n=200) what makes this a
non-additive model. As such, higher-order effects ($S_{T_i}-S_i$) due to
parameters interaction are important and can not be ignored even if $S_i$ is small.
This is the case of Recon that has a high $S_{T_i}$ value and small $S_i$.  In
this example application, the parameters with higher effect values (G1, G2,
Recon) are used in the candidate object segmentation phase, highlighting the
importance of this phase to the results.

\begin{table}[h!]
\begin{center}
\vspace*{-2ex}
\caption{VBD results for the watershed and level set based segmentation workflows.}
\begin{scriptsize}
\subtable[Results for the watershed based segmentation workflow.]{\label{tab:vbd-water}
\begin{tabular}{c c c c c c c}
\hline
\multirow{2}{*}{Parameters}& \multicolumn{2}{c}{n=50}		& \multicolumn{2}{c}{n=100}		& \multicolumn{2}{c}{n=200}		\\ \cline{2-7}
			& Main ($S_i$)	& Total ($S_{T_i}$)	& Main ($S_i$)	& Total ($S_{T_i}$)	& Main ($S_i$)	& Total ($S_{T_i}$)	\\ \hline \hline
	  T2		&-1.25e-05	&1.32e-07		& 2.86e-05   	& 6.36e-08		& 1.67e-03  	& 2.81e-04		\\ \hline
          G1 		& 3.52e-02	&7.57e-02  		&-1.88e-03	& 1.44e-01		& 5.95e-02      & 9.07e-02		\\ \hline
          G2 		& 7.80e-01	&9.46e-01  		& 5.28e-01	& 7.57e-01		& 5.39e-01      & 8.67e-01		\\ \hline
     MinSize 		& 1.73e-02	&3.92e-02  		& 1.67e-02	& 4.13e-02		& 1.34e-02      & 1.58e-02		\\ \hline
     MaxSize 		& 4.76e-03	&2.80e-04  		& 1.65e-03	& 1.70e-03		& 1.29e-04      & 5.39e-04		\\ \hline
   MinSizePl 		&-5.48e-04	&4.80e-02  		& 2.31e-02	& 2.67e-02		& 1.39e-02      & 1.99e-02		\\ \hline
  MinSizeSeg 		& 1.69e-01	&1.95e-01  		& 1.38e-01	& 1.08e-01		& 8.99e-02      & 9.37e-02		\\ \hline
       Recon 		&-2.24e-02	&2.22e-01  		&-2.70e-02	& 3.21e-01		& 2.16e-02      & 2.06e-01		\\ \hline
\rowcolor{maroon!05}Sum & 1.0		&   			& 0.73   	& 			& 0.74   	&			\\ \hline
\end{tabular}}
\subtable[Results for the level set based segmentation workflow.]{\label{tab:vbd-level}
\begin{tabular}{c c c c c c c}
\hline
\multirow{2}{*}{Parameters}& \multicolumn{2}{c}{n=50}		& \multicolumn{2}{c}{n=100}		& \multicolumn{2}{c}{n=200}		\\ \cline{2-7}
			& Main ($S_i$)	& Total ($S_{T_i}$)	& Main ($S_i$)	& Total ($S_{T_i}$)	& Main ($S_i$)	& Total ($S_{T_i}$)	\\ \hline \hline
	  OTSU		&8.91e-01	&7.71e-01		& 9.23e-01  	&9.92e-01 		&9.25e-01   	&9.62e-01 		\\ \hline
          CW 		&7.33e-02	&1.53e-02  		& 1.05e-02      &1.48e-02 		&6.31e-02       &3.61e-02 		\\ \hline
       MinSize 		&1.29e-03	&2.84e-04  		& 1.84e-03      &2.61e-03 		&9.51e-04       &9.46e-04 		\\ \hline
     MsKernel 		&3.15e-02	&2.56e-02  		& 3.09e-02      &2.11e-02 		&1.71e-02       &1.95e-02 		\\ \hline
   LevelSetIt 		&4.88e-03	&2.05e-04  		& 1.03e-03      &2.65e-04 		&2.90e-03       &2.12e-04 		\\ \hline
\rowcolor{maroon!05}Sum &1.0		&   			& 0.96   	&			&0.99    	&			\\ \hline
\end{tabular}}
\end{scriptsize}
\label{tab:vbd}
\end{center}
\end{table}

\begin{figure}[h!]
\begin{center}
        \includegraphics[width=\textwidth]{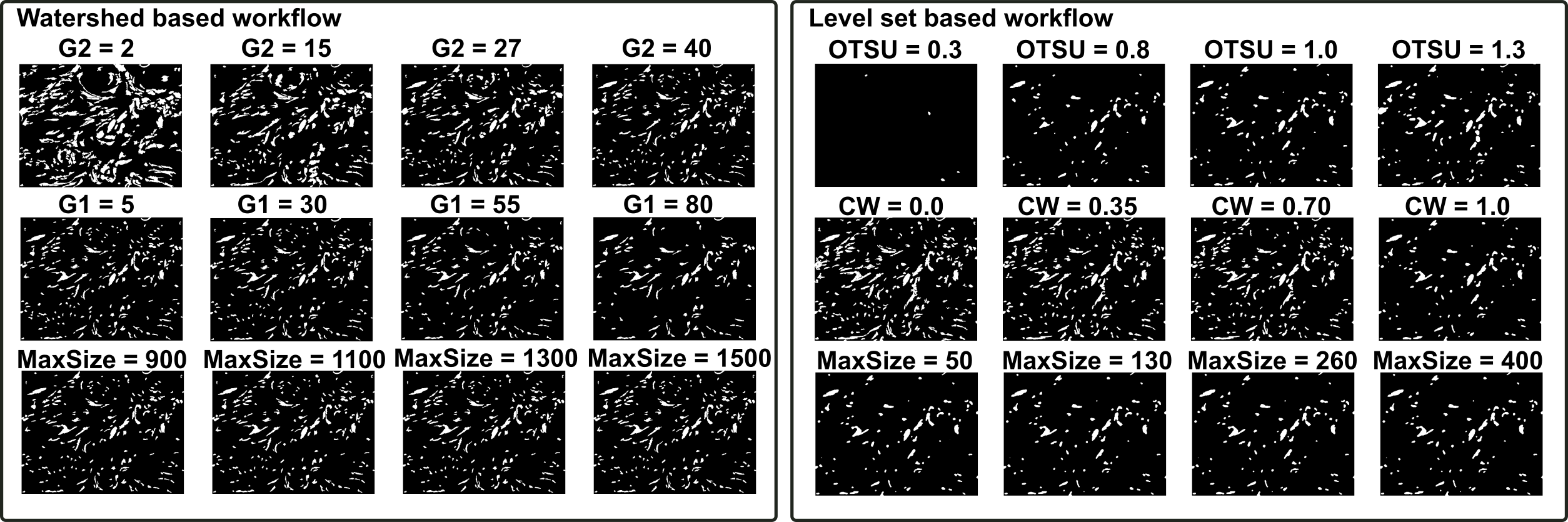}
\caption{Example parameter tuning with output.}
\label{fig:example-output-levelset}
\end{center}
\end{figure}

The VBD results for the level set workflow are presented in
Table~\ref{tab:vbd-level}. In this case, the sum of the main effects is very
close to 1 and the model is additive. The OTSU explains alone most of the
variance on the output results ($S_i$=0.91), but and it has a small higher
effect component resulting from interactions mainly with CW and MsKernel.  The
OTSU parameter is also used in the candidate object identification phase, once
again showing that it is extremely important to reach good segmentation
results. The second most important parameter is the CW, which adapts smoothness
of the nuclei boundaries. Other parameters are less important. Additionally, we
have created a panel with nuclei segmentation results by varying the values of
the two most and the least important parameter from each workflow as
highlighted in the VBD study (See Figure~\ref{fig:example-output-levelset}). As
shown, the amount of variance in the output as a result of the parameter
variations agree with those VBD values computed in the  uncertainty
quantification experiments.

\subsection{Parameter Auto-Tuning}

This section examines the ability of the auto-tuning algorithms in selecting
application parameter values that maximize a metric of interest (Dice and
Jaccard). The experiments where carried out using 15 images manually segmented
by a pathologist, and the experiments were executed for the two use case
workflows. We first performed an experiment in which we tune the application to
maximize the metrics for each image individually to quantify the potential
output quality improvement and to compare tuning algorithms. Further, we
perform a random cross validation that separates images in training and testing
to select a given set of parameters that maximize the quality over a set of
images. The experiment varied parameter values in the segmentation stage as
described in Table~\ref{tab:parameters}.

The results of attained by the auto-tuning algorithms for Dice and Jaccard in
the first experiment set are presented in Table~\ref{tab:tuning-single-image}
-- other metrics are not presented here because they lead to similar result
trends in the comparison of the auto-tuning algorithms. The NM and PRO
algorithms were configured to converge with a maximum of 100 iterations,
whereas the GA was set to evaluated 100 individuals. The GA computed 10
generations with 10 individuals in each of them. We have repeated the GA
experiments 50 times to account for variabilities and the standard deviation
in results is smaller than 1.5\%.

\begin{table}[h!]
\begin{center}
\vspace*{-2ex}
\caption{Comparison of results generated using Default application parameters and those selected by the tuning algorithms NM, PRO, and GA.}
\begin{scriptsize}
\subtable[Results for the watershed based segmentation workflow.]{\label{tab:tuning-single-image-water}
\begin{tabular}{c | c c c c | c c c c}
\hline
\multirow{2}{*}{Image}	& \multicolumn{4}{c}{Dice} & \multicolumn{4}{c}{Jaccard}	\\ \cline{2-9}
	& Default	& NM		&PRO		&GA		& Default& NM		&PRO		&GA		\\ \hline
1	&0.71		&\bf{0.81}	&\bf{0.81}	&0.79   	& 0.55	&\bf{0.68}	&\bf{0.68}	&0.67		\\ \hline
2	&0.75		&\bf{0.83}	&0.81		&0.81   	& 0.60	&\bf{0.70}	&0.68		&0.69		\\ \hline
3	&0.76		&0.80		&\bf{0.81}	&0.80   	& 0.61	&0.67		&\bf{0.68}	&\bf{0.68}	\\ \hline
4	&0.71		&0.84		&0.84		&\bf{0.85}   	& 0.55	&\bf{0.73}	&\bf{0.73}	&\bf{0.73}	\\ \hline
5	&0.71		&0.76		&0.76		&\bf{0.77}   	& 0.55	&\bf{0.62}	&0.61		&\bf{0.62}	\\ \hline
6	&0.82		&\bf{0.85}	&0.84		&0.84   	& 0.70	&\bf{0.73}	&\bf{0.73}	&\bf{0.73}	\\ \hline
7	&0.71		&\bf{0.83}	&\bf{0.83}	&0.82   	& 0.67	&\bf{0.71}	&0.70		&0.70		\\ \hline
8	&0.69		&\bf{0.79}	&\bf{0.79}	&\bf{0.79}   	& 0.53	&\bf{0.65}	&\bf{0.65}	&\bf{0.65}	\\ \hline
9	&0.61		&0.72		&0.72		&\bf{0.73}   	& 0.44	&\bf{0.57}	&\bf{0.57}	&\bf{0.57}	\\ \hline
10	&0.70		&0.78		&\bf{0.79}	&\bf{0.79}   	& 0.53	&0.63		&\bf{0.65}	&\bf{0.65}	\\ \hline
11	&0.57		&0.70		&0.70		&0.72   	& 0.40	&0.54		&0.54		&\bf{0.56}	\\ \hline
12	&0.68		&0.78		&\bf{0.79}	&\bf{0.79}   	& 0.52	&0.64		&\bf{0.66}	&0.65		\\ \hline
13	&0.71		&\bf{0.83}	&\bf{0.83}	&\bf{0.83}   	& 0.60	&\bf{0.71}	&\bf{0.71}	&0.70		\\ \hline
14	&0.79		&\bf{0.84}	&\bf{0.84}	&\bf{0.84}   	& 0.65	&\bf{0.73}	&0.72		&0.72		\\ \hline
15	&0.72		&\bf{0.86}	&\bf{0.86}	&0.85		& 0.66	&\bf{0.76}	&0.75		&0.75		\\ \hline
Sum	&10.65		&12.02		&12.02		&\bf{12.03}	& 8.56	&10.07		&10.06		&\bf{10.07}	\\ \hline
\end{tabular}}
\subtable[Results for the level set based segmentation workflow.]{\label{tab:tuning-single-image-level}
\begin{tabular}{c |c c c c |c c c c}
\hline
\multirow{2}{*}{Image}	& \multicolumn{4}{c}{Dice}		& \multicolumn{4}{c}{Jaccard}	\\ \cline{2-9}
	&Default& NM		&PRO		&GA		&Default	& NM		&PRO		&GA\\ \hline
1	&0.40	&\bf{0.86}	&\bf{0.86}	&0.83   	&0.25 		&\bf{0.76}	&\bf{0.76}	&0.70 \\ \hline
2	&0.10	&0.87		&\bf{0.88}	&0.71   	&0.05 		&\bf{0.77}	&\bf{0.77}	&0.54 \\ \hline
3	&0.04	&0.88		&\bf{0.89}	&0.56  		&0.02 		&0.78		&\bf{0.80}	&0.34 \\ \hline
4	&0.34	&\bf{0.92}	&0.91		&0.85   	&0.20 		&\bf{0.85}	&0.84		&0.59 \\ \hline
5	&0.19	&\bf{0.82}	&\bf{0.82}	&0.75   	&0.10 		&\bf{0.69}	&\bf{0.69}	&0.61 \\ \hline
6	&0.73	&\bf{0.89}	&0.87		&\bf{0.89}   	&0.57 		&0.79		&0.79		&\bf{0.80} \\ \hline
7	&0.69	&\bf{0.88}	&0.87		&\bf{0.88}   	&0.53 		&\bf{0.78}	&0.68		&0.77 \\ \hline
8	&0.77	&\bf{0.86}	&\bf{0.86}	&\bf{0.86}   	&0.63 		&\bf{0.76}	&\bf{0.76}	&0.75 \\ \hline
9	&0.83	&0.86		&0.77		&\bf{0.87}   	&0.71 		&0.74		&0.70		&\bf{0.77} \\ \hline
10	&0.87	&\bf{0.93}	&0.85		&0.89   	&0.77 		&0.76		&0.75		&\bf{0.80} \\ \hline
11	&0.77	&0.72		&0.77		&\bf{0.79}   	&0.63 		&0.61		&0.58		&\bf{0.68} \\ \hline
12	&0.84	&0.84		&0.82		&\bf{0.85}   	&0.72 		&\bf{0.73}	&0.71		&\bf{0.73} \\ \hline
13	&0.90	&0.21		&0.21		&\bf{0.91}   	&\bf{0.82} 	&0.12		&0.12		&\bf{0.82} \\ \hline
14	&0.86	&0.40		&0.41		&\bf{0.88}   	&0.75 		&0.76		&0.63		&\bf{0.78} \\ \hline
15	&0.89	&0.37		&0.37		&\bf{0.91}   	&0.80   	&\bf{0.83}	&0.23		&0.82 \\ \hline
Sum	&9.22	&\bf{11.32}	&11.17		&\bf{11.32}  	&7.56		&\bf{10.72}	&9.81		&10.52 \\ \hline	
\end{tabular}}
\end{scriptsize}
\label{tab:tuning-single-image}
\end{center}
\end{table}

%
%
%

The results presented in Table~\ref{tab:tuning-single-image} show that the
tuning algorithms improved quality of the output (Dice and Jaccard) of the
default algorithms parameters for most input images. In the watershed based
workflow (Table~\ref{tab:tuning-single-image-water}), the average Dice and
Jaccard values are, respectively, 1.14$\times$ and 1.19$\times$ higher than
that of the default parameters, whereas it is 1.22$\times$ and 1.42$\times$
higher in the level set workflow (Table~\ref{tab:tuning-single-image-level}).
The Dice and Jaccard improvements can reach up to 21.4$\times$ and 38.5$\times$
improvement depending on the image used.  The auto-tuning algorithms reached
similar results for both metrics and the watershed case, but higher
differences are observed in the level set workflow tuning in which each
algorithm may attain significantly better results depending on the image used.  As
such, a single tuning algorithm may not be sufficient to maximize the results
quality, but an ensemble of tuning algorithms could be used to select the best
result depending on the input data used.

We further performed a random sub-sampling validation experiment in which we
used 20\% (3) images randomly selected for training the parameter values and
the 80\% (12) remaining images for testing the parameters learned. We have used
the GA algorithm with 100 individuals for tuning and repeated the process 10
times using the Dice metric as an example. For the Watershed workflow, the
learned parameters improved the performance of the default parameters on the
testing data in 1.10$\times$ and 1.13$\times$ on average for Dice and Jaccard
with standard deviation smaller than 1\%.  For the level set based workflow,
the results are even better and the average improvements are 1.29$\times$ and
1.42$\times$ as compared to the default parameter values.. As such, the tuning
could significantly improve the application results selecting more appropriate
parameter values than those picked by a specialist while examining only 100 out
of 21~trillion or 2.8~billion possible parameter combinations.

\subsection{Executing Algorithm/Pipeline Sensitivity Analysis} 
\subsubsection{Performance with Hierarchical Storage}
This section presents the application scalability as the configuration of
hierarchical storage is varied on the Stampede cluster. We evaluated the
storage with 1 level (1L: file system - FS) and 2 levels (2L: RAM+FS), while
the data replacement policy is FIFO or LRU.  We also analyzed performance of
the data locality-aware coarse-grained scheduling (DLAS) as compared to using
the FCFS strategy. A dataset containing 6,113 4K$\times$4K image tiles was
used. 

\begin{figure}[htb!]
\begin{center}
        \includegraphics[width=0.78\textwidth]{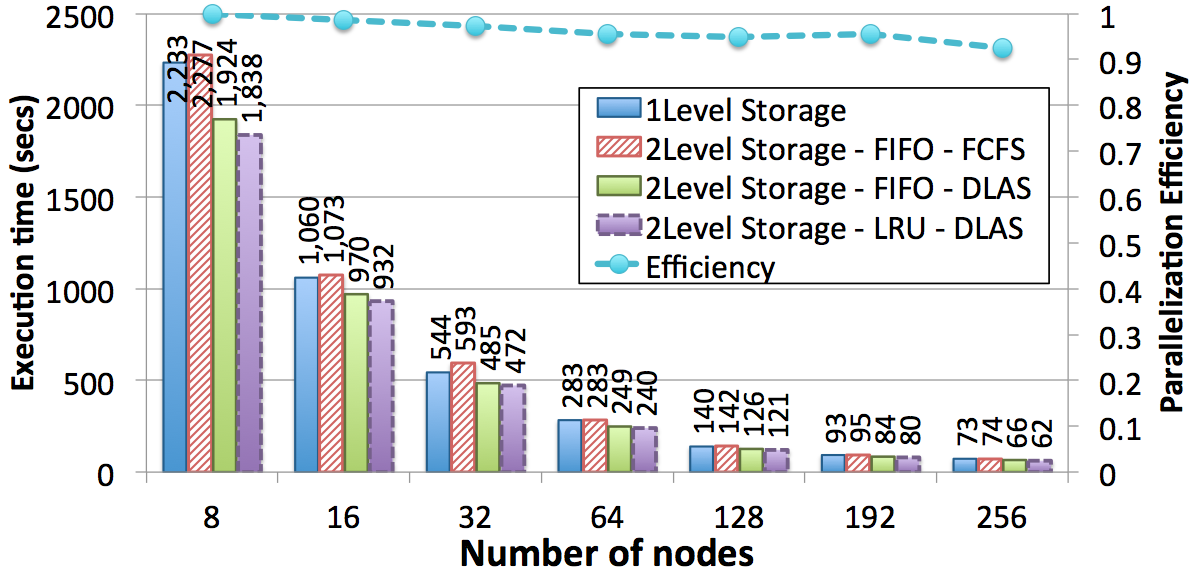}
\caption{Scalability and performance with different storage
and coarse-grained stage scheduling.}
\vspace*{-3ex}
\label{fig:scalability}
\end{center}
\end{figure}

The performance results presented in Figure~\ref{fig:scalability} show that all
versions of the application attained good scalability. The performance of the
configuration with a single storage level is faster than the ``2L FIFO - FCFS".
This is because the 2L FIFO-FCFS setup has an overhead for maintaining an extra
storage level with very low data access hit rate (about 1.5\%) in the first
level storage (RAM). The performance of the ``2L - FIFO - DLAS" configuration
is better than the single level versions for all numbers of nodes (1.11$\times$
on average). This is a result of higher data access hit rate (up to 72\%) in
the first level storage (RAM).  Finally, the ``2L LRU - DLAS"
resulted in the best performance with an average 1.15$\times$ speedup
on the 1L due to improved hit rate (87\%) .

\subsubsection{Performance Impact of Data Reuse} \label{sec:data-reuse}
In this set of experiments, we used a compact representation of the application
workflow graph and varied the number of parameters value sets tested
simultaneously in a run. The 1L storage was used as the baseline, which was
compared to the best storage configuration (LRU + DLAS) with 2 and 3 levels
(3L: RAM+SSD+FS). Only the input parameters of the segmentation stage were
varied. An increase in the number of parameter value sets evaluated in a run
results in a proportional increase in the number of times that output from the
normalization stage is accessed.  Experiments with 1 and 2 storage layers were
executed on both clusters, but no significant difference was observed between
them.  Thus, we only present the results collected on the Eagle cluster, which
is equipped with SSDs and allows for the 3 storage layer configuration.

\begin{table}[h!]
\begin{center}
\vspace*{-2ex}
\caption{Performance of Storage Configurations as data reuse is change by varying \# of parameters
evaluated per run.}
\begin{small}
\begin{tabular}{l c c c c c}
\hline
\multirow{2}{*}{Configuration} 	& \multicolumn{5}{c}{\# of Parameters Evaluated Per Run}	\\ \cline{2-6}
				& 2	& 4	& 8	& 16	& 32	\\ \hline \hline
1L (Baseline)			& 1	& 1	& 1	& 1	& 1	\\ \hline
2L (DLAS+LRU)			& 1.16  & 1.22  & 1.23	& 1.25	& 1.26	\\ \hline
3L (DLAS+LRU)			& 1.3 	& 1.38	& 1.42	& 1.43	& 1.43	\\ \hline
\end{tabular}
\end{small}
\label{tab:datareuse}
\vspace*{-5ex}
\end{center}
\end{table}


Table~\ref{tab:datareuse} presents speedup in the entire application
execution, using the 1L configuration as a baseline, for the storage
configurations with 2 and 3 levels. As is shown, performance gains with the 2L
configuration as compared to the baseline are significant even when a small
number of parameter sets are tested simultaneously. A  speedup of up to $1.26
\times$ is attained when 32 parameter values are tested in the segmentation
stage. Performance improvements, however, are not proportional to the increase
in the number of parameter value sets.  This is because the higher the data
reuse level (\# of parameter values) is, the smaller the data reading time is,
as compared to the entire application computation time. Thus, although data
reading times can be reduced as the number of parameter value sets is
increased, reduction in data read overheads will have little effect in the
application's overall execution time.

The 3L configuration is up to $1.43 \times$ faster than the 1L configuration
due to a reduction on data staging overheads.  In the 3L configuration, data
regions selected to be moved out of RAM are saved in the SSD storage, instead
of being moved directly to the disk-based storage layer.  Since the SSD
storage is larger storage than RAM, data regions can be deleted directly
from SSD after used, minimizing the amount of data staged to the
slower disk-based storage.

\subsubsection{Execution in a Hybrid Setting of CPUs and Co-processors}
In this section, we analyze the performance of the application in a hybrid
setting with CPUs and MICs. The workflow is implemented as a 2-level
hierarchical workflow with the first level being the coarse-grained stages of
normalization, segmentation, feature computation, and comparison. The second
level consists of workflows of operations that implement each of the stages as
presented in Section~\ref{sec:intro}. Five versions of the hybrid setup were
evaluated: (1)~CPU-only uses all CPU cores; (2)~MIC-only uses only the
co-processors; (3)~CPU-MIC FCFS uses the CPU cores and co-processor with
distribution of tasks among processors using FCFS (First-Come, First-Served);
(4)~CPU-MIC HEFT uses the CPU cores and co-processor with distribution of tasks
among processors using HEFT (Heterogeneous Earliest Finish Time); (5)~CPU-MIC
PATS uses the CPU cores and co-processors with the PATS scheduler for task
scheduling.

\begin{figure}[h!]
\begin{center}
        \includegraphics[width=0.78\textwidth]{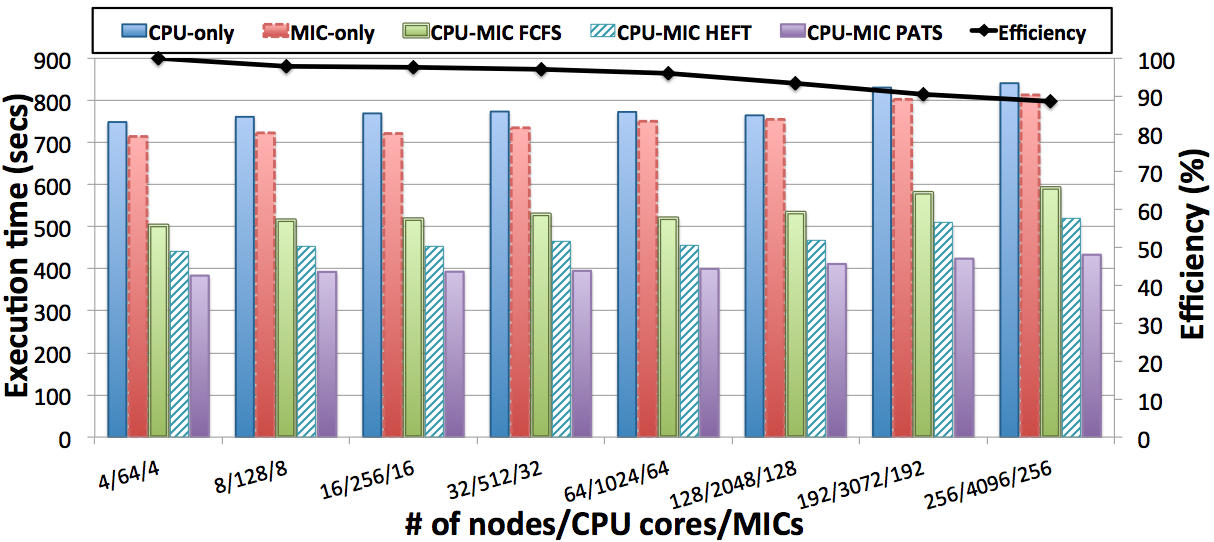}
\caption{Task scheduling strategies in a weak scaling evaluation.}
\label{fig:sched}
\end{center}
\end{figure}

The evaluation was performed in a weak scaling experiment in which dataset
size and computation nodes are increased proportionately. The experiment dataset 
contains up to 136,568 4K$\times$4K image tiles (6.5TB of uncompressed
data) when the number of nodes is 256. The results in
Figure~\ref{fig:sched} show that all versions of the application scaled well and 
that cooperative execution with the hybrid configuration led to significant performance
gains. Moreover, the use of our PATS scheduler improved the performance on top
of FCFS and HEFT on average by about 1.32$\times$ and 1.2$\times$, respectively.
Performance gains from PATS result from the better ability of PATS in taking into account 
heterogeneity in performances (speedups) of different tasks when assigning tasks to 
processors. This observation corroborates with results from our earlier 
work~\cite{Teodoro:2014:CPA:2650283.2650645}, which used a workflow of segmentation 
and feature computation stages only and compared PATS against FCFS. 

\subsubsection{Simultaneous Parameter Evaluation} \label{sec:opt-search}
This section analyzes the impact of performing simultaneous multi-parameter
evaluation available with PRO and GA (see Section~\ref{sec:auto-parallel}). The
results were collected using Set~1 of the reference mask from the previous
section.  In these experiments, we varied the number of parameter sets
evaluated simultaneously in each iteration of the tuning process. In each
iteration, the compact graph representation is built to efficiently execute
parameter evaluation. Computing resources are fixed across the experiments.

\begin{table}[h!]
\begin{center}
\vspace*{-1ex}
\caption{Speedups due simultaneous parameter evaluation with application configurations C1 and C2.}
\begin{scriptsize}
\begin{tabular}{l l l l l l l l l}
\hline
 			& 		& \multicolumn{7}{c}{\# of param. sets evaluated per iteration}	\\ \cline{3-9} \cline{3-9}
			& 		& 2	& 3	& 4	& 5	& 6	& 7	& 8		\\ \hline \hline
\multirow{2}{*}{C1}	& Observed	& 1.22	& 1.40	& 1.47	& 1.54	& 1.58	& 1.58	& 1.60		\\ \cline{3-9}
			& Up. Limit	& 1.28 	& 1.42	& 1.50	& 1.55	& 1.58	& 1.61	& 1.63		\\ \hline
\multirow{2}{*}{C2}	& Observed	& 1.30  & 1.57  & 1.68  & 1.78  & 1.84	& 1.89	& 1.91		\\ \cline{3-9}
			& Up. Limit	& 1.38 	& 1.59	& 1.72	& 1.81	& 1.87	& 1.92	& 1.95		\\ \hline
\end{tabular}
\end{scriptsize}
\label{tab:par-tuning}
\end{center}
\end{table}

We used two application configurations, C1 and C2, that differ in the
computation cost of the segmentation stage. The goal is to show that
performance gains with an optimized parameter evaluation vary according to the
characteristics of the application.  The version C1 of the application is
implemented as in the previous section with the normalization, segmentation,
and comparison stages.  The cost of the normalization stage, which may be
reused, in C1 consists of about 45\% of the entire execution time. The version
C2 has the same workflow of stages but the computation cost of the segmentation
stage is reduced. As a result, the normalization stage takes about 55\% of the
entire execution time.

Table~\ref{tab:par-tuning} presents speedups as the number of parameters tested
simultaneously in an iteration of the tuning is increased for both
configurations (C1 and C2). The speedup values are calculated using as a
baseline the version in which a single parameter set is evaluated per
iteration. As is shown, the compact representation and
simultaneous parameter evaluation led to performance improvements of about
1.63$\times$ and 1.95$\times$ in C1 and C2, respectively. The table also
represents the maximum speedup that could be attained by simultaneous parameter
testing (``Up.  Limit") in each case. This value was calculated by measuring
the contribution of each workflow stage to the entire execution time and
removing possible common computation paths in each case. The performance
improvements attained in practice are very close the upper limit in most of the
cases. As is shown in the table, the performance improvements with the
simultaneous evaluation of parameter sets increase slower when the number of
parameter sets evaluated per iteration is higher. This is because the larger
the number of parameters tested simultaneously is, the smaller is the amortized
cost of common paths that could be reused, since not-common paths start
dominating the application execution. This is a good characteristic of the
method, because high performance gains can be attained without the need of a
very large number of parameter sets for simultaneous evaluation.

%

\section{Conclusions} \label{sec:conclusions}
We propose and demonstrate a set of runtime optimizations for efficient
execution of algorithm sensitivity analysis applications. We show that
auto-tuning designed for application execution performance can be employed for
improving analysis results. The cost of parameter space search can be reduced
by simultaneously evaluating multiple parameter values on a cluster system
while eliminating duplicate computations. We successfully demonstrate the
impact of the proposed optimizations by tuning a complex cancer image analysis
application using a large-scale cluster system and large datasets. We argue
that the use of the proposed runtime optimizations coupled with auto-tuning
algorithms can enable systematic, comparative study of analysis pipelines and
improve analysis results when large datasets need to be analyzed. 
As a future work, we intend to evaluate other auto-tuning algorithms.
We will also implement other comparative analysis
tasks and optimizations to reuse computation in fine-grain tasks.
Another future extension of our work will be its
integration with visual parameter optimization tools and
interfaces~\cite{Pretorius2015,journals/tvcg/SchultzK13} to allow for
post-tuning a finer gain and visual analyses of collected results. \\

\vspace*{-3mm}
\noindent {\bf Acknowledgments}.
This work was supported in part by 1U24CA180924-01A1 from
the NCI, R01LM011119-01 and R01LM009239 from the
NLM, CNPq, and NIH K25CA181503. This research used resources of the 
XSEDE Science Gateways program under grant TG-ASC130023.

\nocite{10.1109/ICPP.2008.72,kong2015automated,stegmaier2016real}

\bibliography{george}

\end{document}